\documentclass[runningheads]{llncs}
\usepackage{graphicx}
\usepackage{amsmath,amssymb}
\usepackage{mathrsfs}
\usepackage{rotating}
\usepackage{ctable}
\usepackage{bm}
\usepackage{color}
\usepackage{multirow}
\usepackage{floatrow}
\usepackage{blindtext}

\newfloatcommand{capbtabbox}{table}[][\FBwidth]

\usepackage{eso-pic}
\usepackage{blindtext}

\begin{document}
\title{LIFE: A Generalizable Autodidactic Pipeline for 3D OCT-A Vessel Segmentation}

\author{Dewei Hu\inst{1}\and
Can Cui\inst{1}\and
Hao Li\inst{1}\and
Kathleen E. Larson\inst{2}\and
Yuankai K. Tao\inst{2}\and
Ipek Oguz\inst{1}}
\authorrunning{D. Hu et al.}
\institute{Vanderbilt University, Dept. of Electrical Engineering and Computer Science
\and
Vanderbilt University, Dept. of Biomedical Engineering, Nashville, TN, USA}
%
%
%
%
\titlerunning{LIFE for 3D OCT-A segmentation}
%
%
\maketitle              
\begin{abstract}
Optical coherence tomography (OCT) is a non-invasive imaging technique widely used for ophthalmology. It can be extended to OCT angiography (OCT-A), which reveals the retinal vasculature with improved contrast. Recent deep learning algorithms produced promising vascular segmentation results; however, 3D  retinal vessel segmentation remains difficult due to the lack of manually annotated training data. We propose a learning-based method that is only supervised by a self-synthesized modality named local intensity fusion (LIF). LIF is a capillary-enhanced volume computed directly from the input OCT-A. We then construct the local intensity fusion encoder (LIFE) to map a given OCT-A volume and its LIF counterpart to a shared latent space. The latent space of LIFE has the same dimensions as the input data and it contains features common to both modalities. By binarizing this latent space, we obtain a volumetric vessel segmentation. Our method is evaluated in a human fovea OCT-A and three zebrafish OCT-A volumes with manual labels. It yields a Dice score of 0.7736 on human data and $0.8594\pm 0.0275$ on zebrafish data, a dramatic improvement over existing unsupervised  algorithms.

\end{abstract}
\section{Introduction}
Optical coherence tomography (OCT) is a non-invasive imaging technique that provides high-resolution volumetric visualization of the retina~\cite{li2017statistical}. However, it offers poor contrast between vessels and nerve tissue layers~\cite{gao2016optical}. This  can be overcome by decoupling the dynamic blood flow within vessels from stationary nerve tissue by 
decorrelating multiple cross-sectional images (B-scans) taken at the same spatial location. 
By computing the variance of these repeated B-scans, we obtain an OCT angiography (OCT-A) volume that has better visualization of retinal vasculature than traditional OCT~\cite{jia2012split}. In contrast to other techniques such as fluorescein angiography (FA), OCT-A is advantageous because it both provides depth-resolved information in 3D and is free of risks related to dye leakage or potential allergic reaction~\cite{gao2016optical}. OCT-A is popular for studying various retinal pathologies~\cite{burke2017application,ishibazawa2015optical}.
Recent usage of the vascular plexus density as a disease severity indicator~\cite{hollo2018comparison} highlights the need for vessel segmentation in OCT-A. 

Unlike magnetic resonance angiography (MRA) and computed tomography angiography (CTA), OCT-A suffers from severe speckle noise, which induces poor contrast and vessel discontinuity. Consequently, unsupervised vessel segmentation approaches~\cite{bozkurt2020texture,zhao2018vascular,993126,Lorigo:2001jv,Vasilevskiy01fluxmaximizing} developed for other modalities do not translate well to OCT-A. Denoising OCT/OCT-A images has thus been an active topic of research~\cite{oguz2020self,hu2020retinal,devalla2019deep}. The noise is compounded in OCT-A due to the unpredictable patterns of blood flow as well as artifacts caused by residual registration errors, which lead to insufficient suppression of stationary tissue. This severe noise level, coupled with the intricate detail of the retinal capillaries, leads to a fundamental roadblock to 3D segmentation of the retinal blood vessels: the task is too challenging for unsupervised methods, and yet, obtaining manual segmentations to train supervised models is prohibitively expensive. For instance, a single patch capturing only about 5\% of the whole fovea {\bf (Fig.~\ref{fig:GT}f)} took approximately 30 hours to manually segment. The large inter-subject  variability and the vast inter-rater variability which is inevitable in such a detailed task make the creation of a suitably large manual training dataset intractable.

As a workaround, retinal vessel segmentation attempts have been largely limited to 2D images with better SNR, such as the depth-projection of the OCT-A~\cite{giarratano2019automated}. 
This only produces a single 2D segmentation out of a whole 3D volume, evidently sacrificing the 3D depth information. Similar approaches to segment inherently 2D data such as fundus images have also been reported~\cite{lahiri2016deep}. Recently, Liu et al.~\cite{liu2020variational} proposed an unsupervised 2D vessel segmentation method using two registered modalities from two different imaging devices. Unfortunately, multiple scans of a single subject are not typically available in practice. Further, the extension to 3D can be problematic due to inaccurate volumetric registration between modalities. Zhang et al.~proposed the optimal oriented flux~\cite{law2008three} (OOF) for 3D OCT-A segmentation \cite{8879535}, but, as their focus is shape analysis, neither a detailed discussion nor any numerical evaluation on segmentation are provided.

\begin{figure}[t]
    \centering
    \includegraphics[width=.95\linewidth]{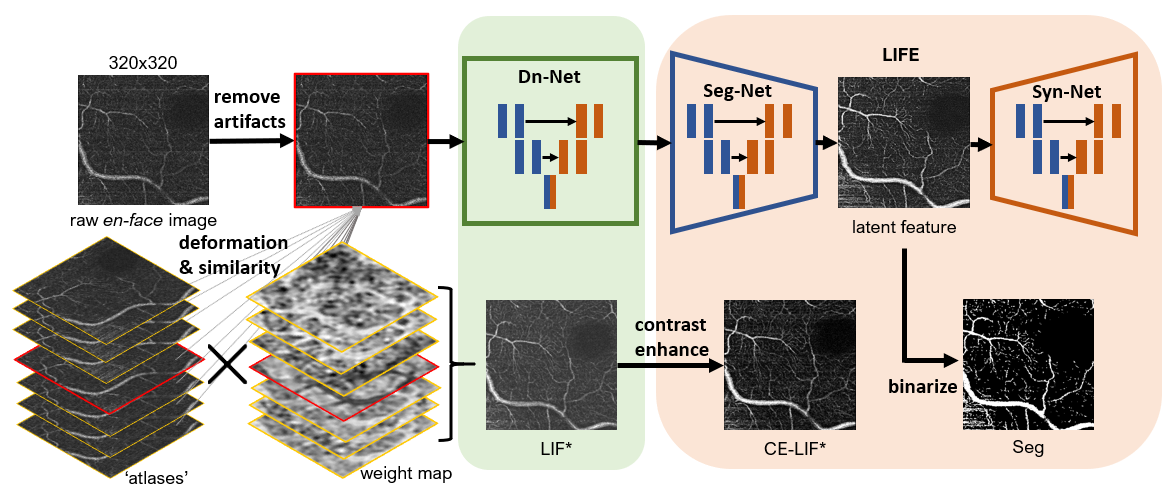}
    \caption{Overall pipeline. * indicates LIF and CE-LIF provide supervision for Dn-Net and LIFE in training process respectively.}
    \label{fig:pipeline}
\end{figure}

We propose the local intensity fusion encoder (LIFE), a self-supervised method to segment 3D retinal vasculature from OCT-A. LIFE requires neither manual delineation nor multiple acquisition devices. To our best knowledge, it is the first label-free learning method with quantitative validation of 3D OCT-A vessel segmentation. {\bf Fig.~\ref{fig:pipeline}} summarizes the pipeline. Our {\bf novel contributions} are: 
\begin{itemize}
    \item An \emph{en-face} denoising method for OCT-A images via local intensity fusion (LIF) as a new modality (Sec.~\ref{sec:LIF})
    \item A variational auto-encoder network, the local intensity fusion encoder (LIFE), that considers the original OCT-A images and the LIF modality to estimate the latent space which contains the retinal vasculature (Sec.~\ref{sec:life})
    \item Quantitative and qualitative evaluation on human and zebrafish data (Sec.~\ref{sec:eval})
\end{itemize}

\section{Methods}

\subsection{Local intensity fusion: LIF}
\label{sec:LIF}


Small capillaries have low intensity in OCT-A since they have slower blood flow, and are therefore hard to distinguish from the ubiquitous speckle noise. We exploit the similarity of vasculature between consecutive \emph{en-face} OCT-A slices to improve the image quality. This local intensity fusion (LIF) technique derives from the Joint Label Fusion~\cite{wang2012multi} and related synthesis methods \cite{fleishman2017joint,10.1117/12.2550009,oguz2020self}. 

\iftrue
Joint label fusion (JLF)~\cite{wang2012multi} is a well-known multi-atlas label fusion method for segmentation. In JLF, a library of $K$ atlases with known segmentations $(\bm{X}_k,\bm{S}_k)$ is deformably registered to the target image $\bm{Y}$ to obtain $(\hat{\bm{X}_k},\hat{\bm{S}_k})$. Locally varying weight maps are computed for each atlas based on the local residual registration error between $\bm{Y}$ and $\hat{\bm{X}_k}$. The weighted sum of the $\hat{\bm{S}}_k$ provides the consensus segmentation $\bm{S}$ on the target image. 

JLF has been extended to joint intensity fusion (JIF), an image synthesis method that does not require atlas segmentations. JIF has been used for lesion in-painting \cite{fleishman2017joint} and cross-modality synthesis \cite{10.1117/12.2550009}. Here, we propose a JIF variant, LIF, performing fusion between the 2D \emph{en-face} slices of a 3D OCT-A volume. 

Instead of an external group of atlases, for each 2D \emph{en-face} slice $\bm{X}$ of a 3D OCT-A volume, adjacent slices within an R-neighborhood $\{\bm{X}_{-R},\hdots,\bm{X}_{+R}\}$ are regarded as our group of 'atlases' for $\bm{X}$. Note that the atlas $X_0$ is the target $X$ itself, represented as the image with a red rim in {\bf Fig.~\ref{fig:pipeline}}. We perform registration using the \emph{greedy} software~\cite{yushkevich2016fast}. While closely related, we note that the self-fusion method reported in~\cite{oguz2020self} for tissue layer enhancement is not suitable for vessel enhancement as it tends to substantially blur and distort blood vessels~\cite{hu2020retinal}.
\fi
\begin{figure}[t]
    \centering
    \begin{tabular}{ccc}
        \includegraphics[width=0.3\linewidth]{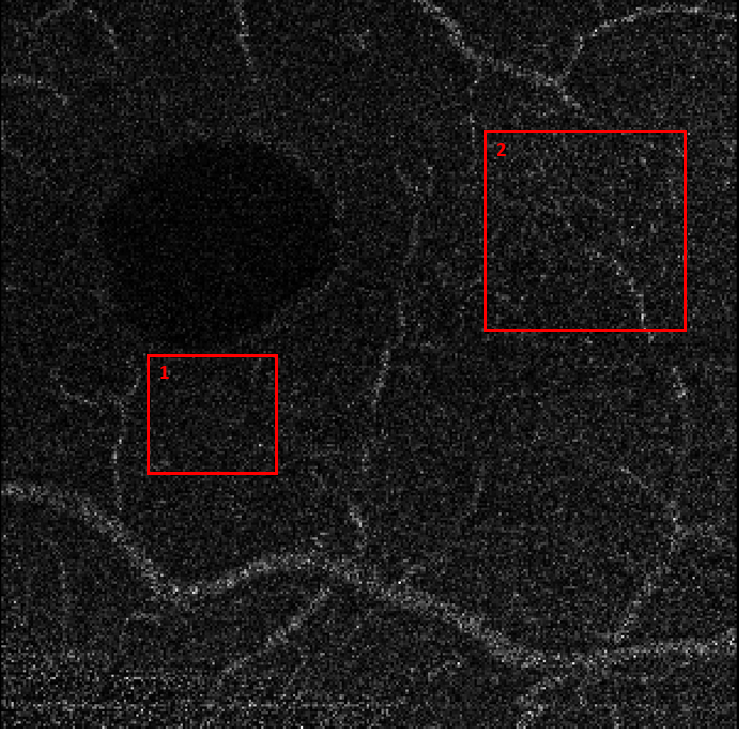} &
        \includegraphics[width=0.3\linewidth]{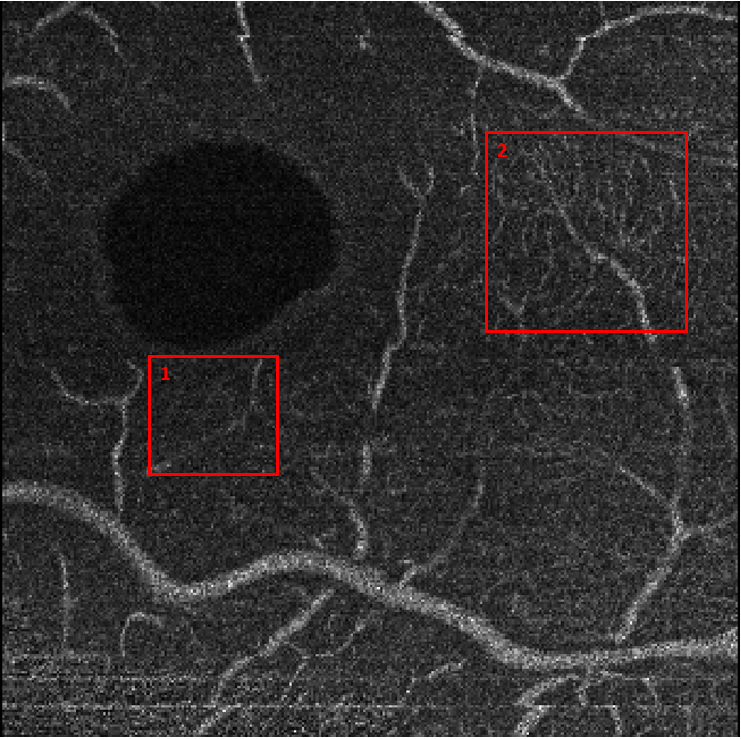} &
        \includegraphics[width=0.3\linewidth]{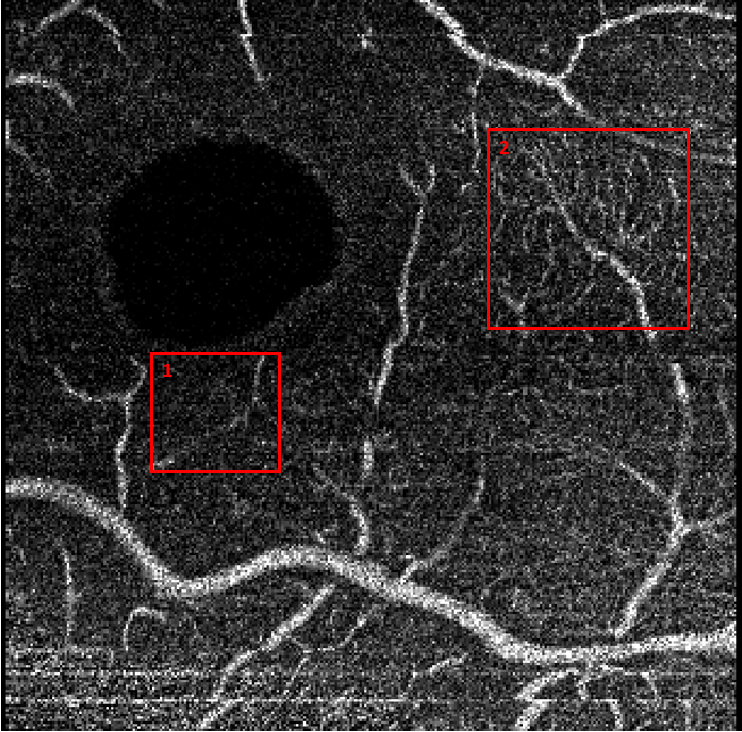}\\
        (a) original & (b) LIF & (c) CE-LIF
    \end{tabular}
    \caption{Modalities of \emph{en-face} OCT-A. {\bf Large red box} highlights improvement in capillary visibility. {\bf Small red box} points out a phantom vessel.}
    \label{fig:en-face}
\end{figure}

Similar to a 1-D Gaussian filter along the depth axis, LIF has a blurring effect that improves the homogeneity of vessels without dilating their thickness in the \emph{en-face} image. Further, it can also smooth the speckle noise in the background while raising the overall intensity level, as shown in {\bf Fig.~\ref{fig:en-face}a/\ref{fig:en-face}b}. In order to make vessels stand out better, we introduce the contrast enhanced local intensity fusion (CE-LIF)\footnote{https://pillow.readthedocs.io/en/stable/reference/ImageEnhance.html} in {\bf Fig.~\ref{fig:en-face}c}. However, intensity fusion of \emph{en-face} images sacrifices the accuracy of vessel diameter in the depth direction. Specifically, some vessels existing exclusively in neighboring images are inadvertently projected on the target slice. For example, the small red box in {\bf Fig.~\ref{fig:en-face}} highlights a phantom vessel caused by incorrect fusion. As a result, LIF and CE-LIF are not appropriate for direct use in application, in spite of the desirable improvement they offer in visibility of capillaries (e.g., large red box). In the following section, we propose a novel method that allows us to leverage LIF as an auxiliary modality for feature extraction during which these excessive projections will be filtered out.    

\subsection{Cross-modality feature extraction: LIFE}
\label{sec:life}
Liu et al.~\cite{liu2020variational} introduced an important concept for unsupervised feature extraction. Two depth-projected 2D OCT-A images, $\bm{M}_1$ and $\bm{M}_2$, are acquired using different devices on the same retina. If they are well aligned, then aside from noise and difference in style, the majority of the anatomical structure would be the same. A variational autoencoder (VAE) is set as a pix2pix translator from $\bm{M}_1$ to $\bm{M}_2$ in which the latent space $\bm{L}_{12}$ keeps full resolution. 
\begin{equation}
    \bm{L}_{12} = f_e(\bm{M}_1)
\qquad \mathrm{and} \qquad
    \bm{M}_2^{'} = f_d(\bm{L}_{12})
\end{equation}
If $\bm{M}_2$ is well reconstructed ($\bm{M}_2^{'}\approx\bm{M}_2$), then the latent feature map $\bm{L}_{12}$ can be regarded as the common features between $\bm{M}_1$ and $\bm{M}_2$, namely, vasculature.  The encoder $f_e$ is considered a segmentation network (Seg-Net) and the decoder $f_d$ a synthesis network (Syn-Net). 

Unfortunately, this method has several drawbacks in practice. Imaging the same retina with different devices is rarely possible even in research settings and unrealistic in clinical practice. Furthermore, the 3D extension does not appear straightforward due to the differences in image spacing between OCT devices and the difficulty of volumetric registration in these very noisy images. In contrast, we propose to use a single OCT-A volume $X$ and its LIF, $X_{LIF}$, as the two modalities. This removes the need for multiple devices or registration, and allows us to produce a 3D segmentation by operating on individual \emph{en-face} OCT-A slices rather than a single depth-projection image. We call the new translator network local intensity fusion encoder (LIFE). 

\begin{figure}[t]
    \centering
    \includegraphics[width=.9\linewidth]{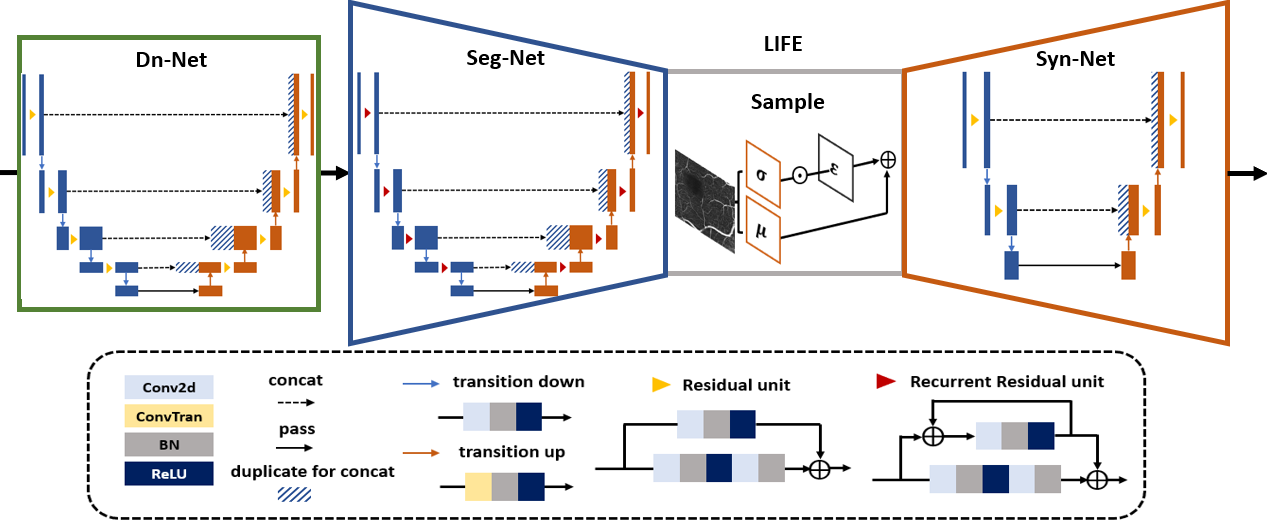}
    \caption{Network architecture}
    \label{fig:VAE}
\end{figure}

Fig.~\ref{fig:VAE} shows the network architecture.
To reduce the influence of speckle noise, we train a residual U-Net as a denoising network (Dn-Net), supervised by LIF. For the encoder, we implement a more complex model (R2U-Net)~\cite{alom2018recurrent} than Liu et al.~\cite{liu2020variational}, supervised by CE-LIF. As the decoder we use a shallow, residual U-Net to balance computational power and segmentation performance. The reparameterization trick enables gradient back propagation when sampling is involved in a deep network~\cite{kingma2013auto}. This sampling is achieved by $
    \bm{S} = \bm{\mu}+\bm{\sigma} \cdot \bm{\epsilon}
$,
where $\bm{\epsilon}\in \mathcal{N}(0,1)$, and $\bm{\mu}$ and $\bm{\sigma}$ are mean and standard deviation of the latent space.  The intensity ranges of all images are normalized to $[0,255]$. To introduce some blurring effect, both $L_1$ and $L_2$ norm are added to the VAE loss function: 
\begin{equation}
\label{loss}
Loss = a \sum_{i,j} |\bm{Y}(i,j)-\bm{Y}^{'}(i,j)|+\frac{b}{N} \sum_{i,j} (\bm{Y}(i,j)-\bm{Y}^{'}(i,j))^2
\end{equation}
where (i,j) are pixel coordinates, N is the number of pixels, $\bf{Y}$ is CE-LIF and $\bf{Y}^{'}$ is the output of Syn-Net. $a=1$ and $b=0.05$ are hyperparameters. Eq.~\ref{loss} is also used as the loss for the Dn-Net, with the LIF image as $\bf{Y}$, $a=1$ and $b=0.01$.

As discussed above, LIF enhances the appearance of blood vessels but also introduces phantom vessels because of fusion. The set of input vessel features $\mathcal{V}$ in $X$ will thus be a subset of $\mathcal{V}_{LIF}$ in $X_{LIF}$. Because LIFE works to extract $\mathcal{V}\cap\mathcal{V}_{LIF}$, the phantom features that exist only in $\mathcal{V}_{LIF}$ will be cancelled out as long as the model is properly trained without suffering from overfitting.

\subsection{Experimental details}

\begin{figure}[t]
    \centering
    \begin{tabular}{cc|c}
        \begin{tabular}{c}
        \includegraphics[width=.3\linewidth]{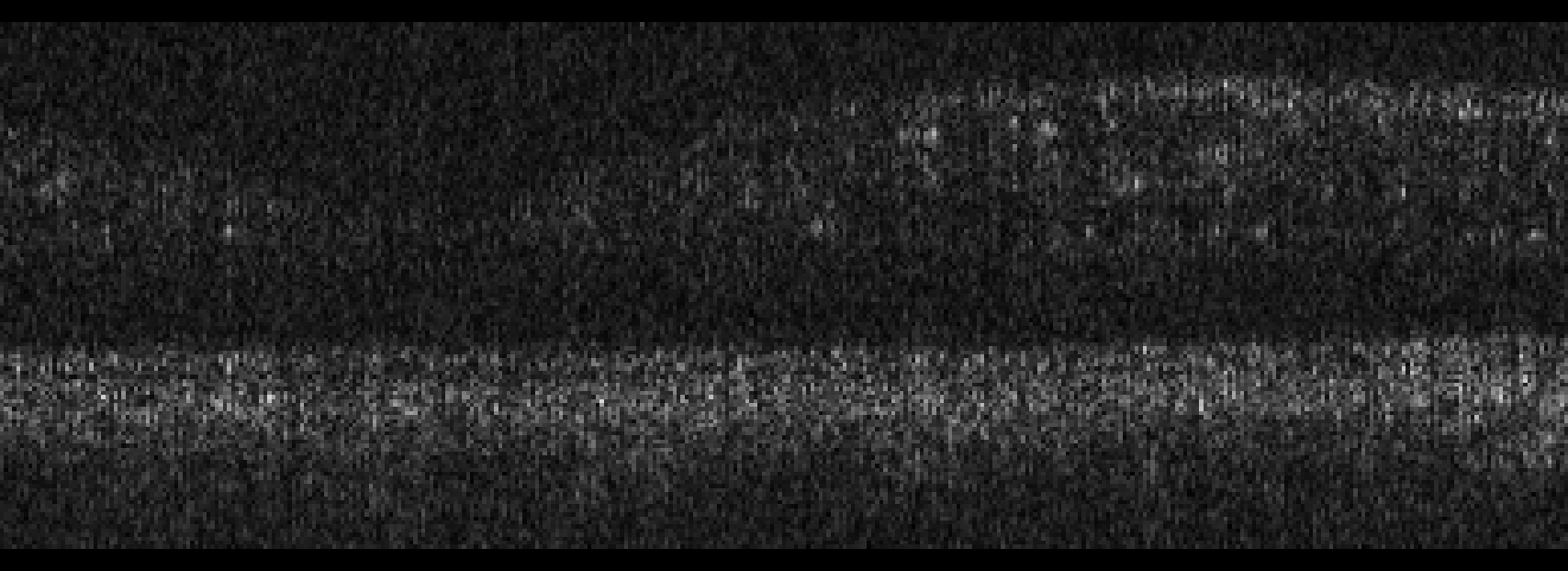}\\
        (a) B-scan without artifact \\
        \includegraphics[width=.3\linewidth]{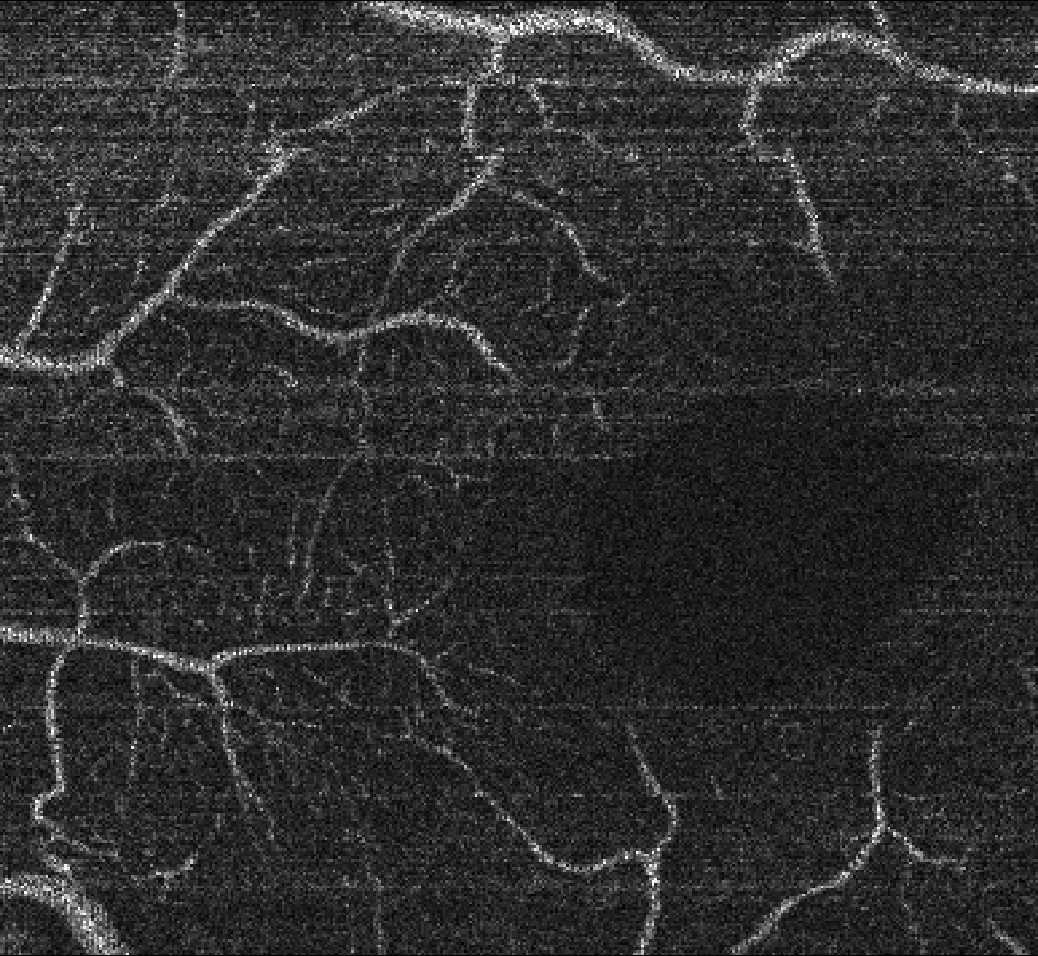}\\
        (c) original \emph{en-face} image
        \end{tabular} &
        \begin{tabular}{c}
         \includegraphics[width=.3\linewidth]{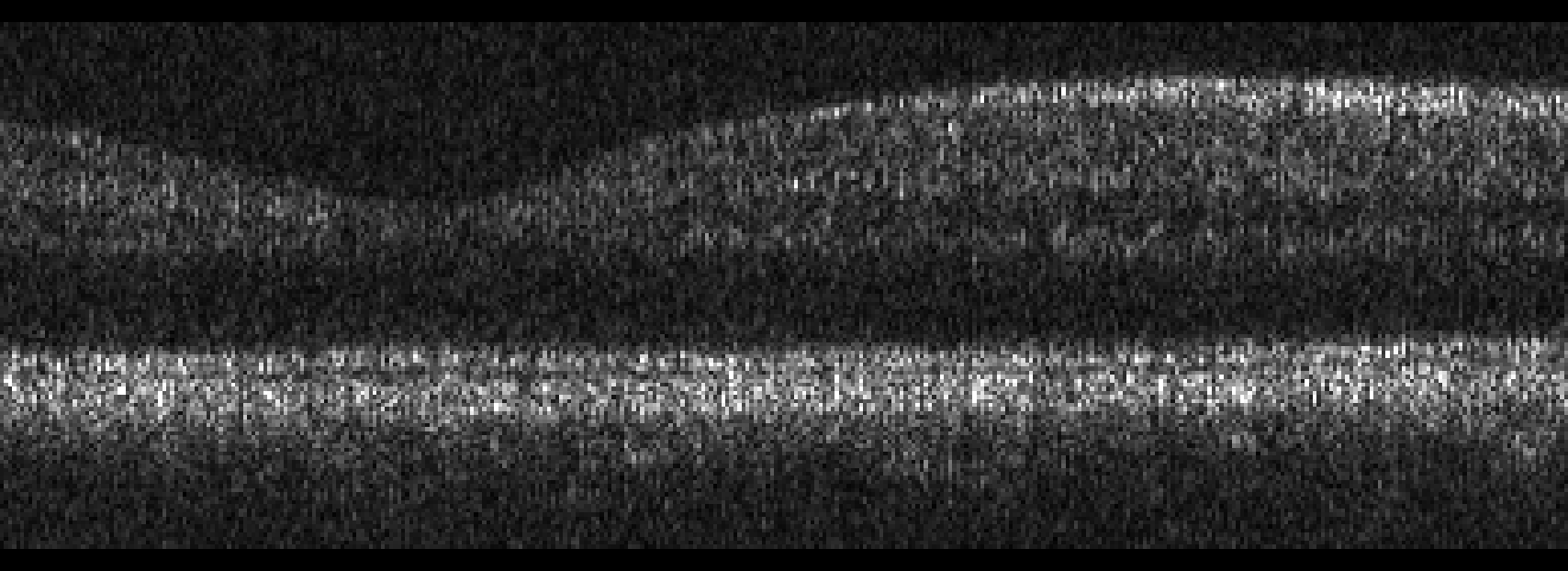}\\
        (b) B-scan with artifact\\
        \includegraphics[width=.3\linewidth]{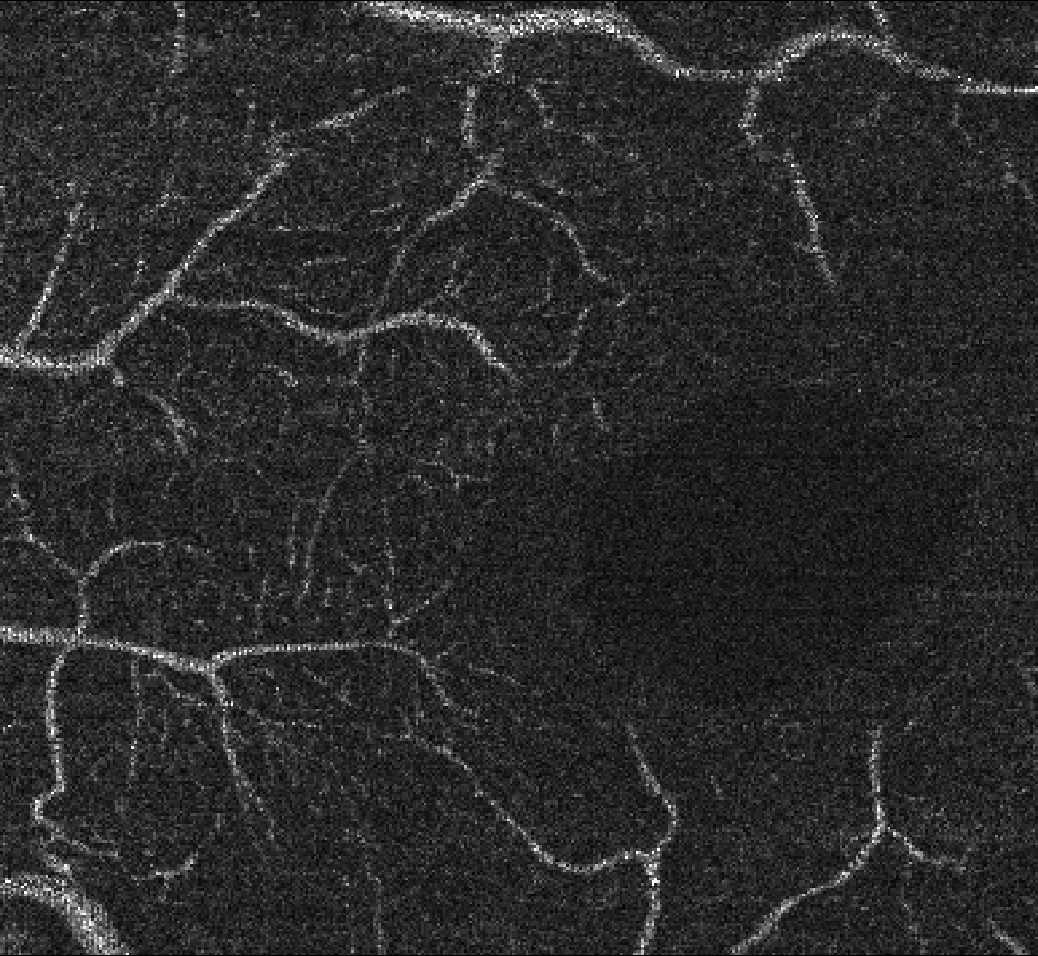}\\
        (d) cleaned \emph{en-face} image 
         \end{tabular} &
         \begin{tabular}{c}
         \includegraphics[width=.2\linewidth]{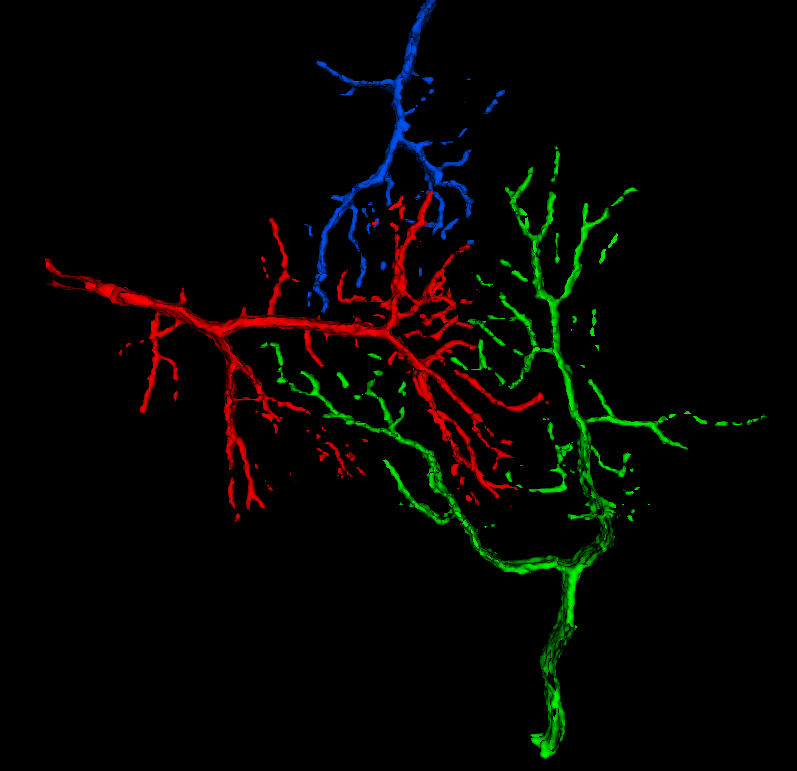}\\
        (e) whole plexus\\
         \includegraphics[width=.2\linewidth]{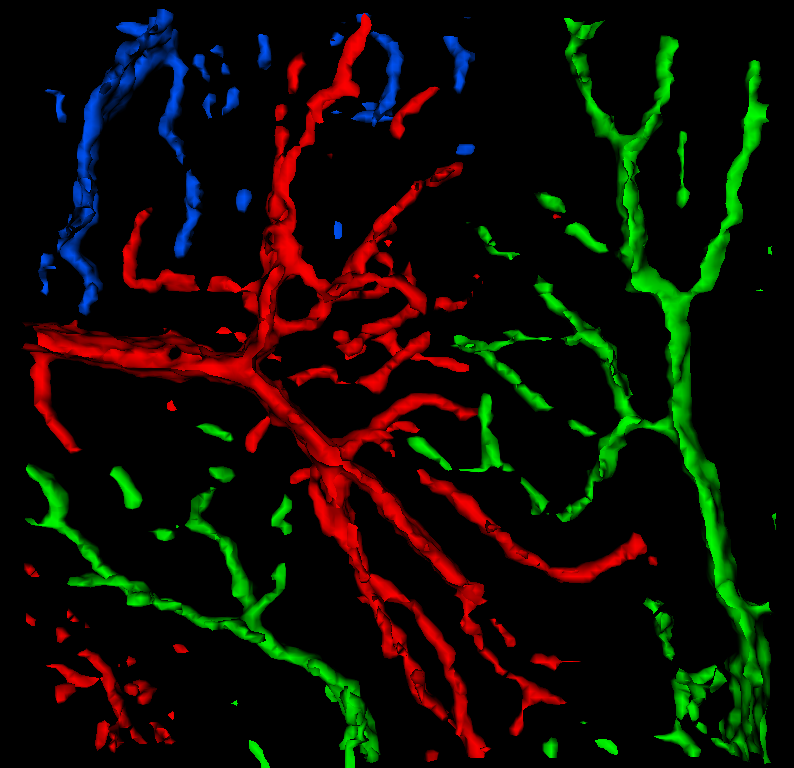}\\
         (f) ROI 
         \end{tabular}
    \end{tabular}
    \caption{{\bf (a-d)} Horizontal motion artifacts and their removal. {\bf (e,f)} Three manually labelled vessel plexuses. Only the branches contained within the ROI were fully segmented; any branches outside the ROI and all other trees were omitted for brevity.}
    \label{fig:GT}
    \label{fig:art}
\end{figure}

{\bf Preprocessing for motion artifact removal.}
Decorrelation allows OCT-A to emphasize vessels while other tissue types get suppressed {\bf (Fig.~\ref{fig:art}a)}. However, this requires the repeated OCT B-scans to be precisely aligned. Any registration errors cause motion artifacts, such that stationary tissue is not properly suppressed {\bf (Fig.~\ref{fig:art}b)}. These appear as horizontal artifacts in \emph{en-face} images {\bf (Fig.~\ref{fig:art}c)}. We remove these artifacts by matching the histogram of the artifact B-scan to its closest well-decorrelated neighbor {\bf (Fig.~\ref{fig:art}d)}. 

{\bf Binarization.}
To binarize the latent space $L_{12}$ estimated by LIFE, we apply the $2^{nd}$ Perona-Malik diffusion equation~\cite{perona1990scale} followed by the global Otsu threshold~\cite{otsu1979threshold}. Any islands smaller than 30 voxels are removed.

{\bf Dataset.} 
The OCT volumes were acquired with $2560 \times 500 \times 400 \times 4$ pix. (spectral $\times$ lines $\times$ frames $\times$ repeated frames)~\cite{malone2019handheld,el2018spectrally}. OCT-A is performed on  motion-corrected~\cite{guizar2008efficient} OCT volumes using singular value decomposition. We manually crop the volume to only retain the depth slices that contain most of the vessels near the fovea, between the ganglion cell layer (GCL) and inner plexiform layer (IPL). Three fovea volumes are used for training and one for testing. As the number of slices between GCL and IPL is limited, we aggressively augment the dataset by randomly cropping and flipping 10 windows of size $[320,320]$ for each \emph{en-face} image. To evaluate on vessels differing in size, we labeled 3 interacting plexus near the fovea, displayed in {\bf Fig.~\ref{fig:GT}e}. A smaller ROI ({$120\times 120\times 17$}) cropped in the center {\bf (Fig.~\ref{fig:GT}f)} is used for numerical evaluation. 

To further evaluate the method, we train and test our model on OCT-A of zebrafish eyes, which have a simple vessel structure ideal for easy manual labeling. This also allows us to test the generalizability of our method to images from different species. Furthermore, the fish dataset contains stronger speckle noise than the human data, which allows us to test the robustness of the method to  high noise. 3 volumes ({$480\times 480\times 25 \times 5$ each}) are labeled for testing and 5 volumes are used for training. All manual labelling is done on ITKSnap~\cite{py06nimg}.

{\bf Baseline methods.}
Due to the lack of labeled data, no supervised learning method is applicable. Similar to our approach that follows the enhance + binarize pattern, we apply Frangi's multi-scale vesselness filter~\cite{frangi1998multiscale} and optimally oriented flux (OOF)~\cite{8879535,law2008three} respectively to enhance the artifact-removed original image, then use the same binarization steps described above. We also present results using Otsu thresholding and k-means clustering.

{\bf Implementation details.}
All networks are trained on an NVIDIA RTX 2080TI 11GB GPU for 50 epochs with batch size set to 2.  For the first 3 epochs, the entire network uses the same Adam optimizer with learning rate of 0.001. After that, LIFE and decoder are separately optimized with starting learning rates of 0.002 and 0.0001 respectively in order to distribute more workload on the LIFE. Both networks decay every 3 epochs with at a rate of 0.5. 

\begin{figure}[t]
    \centering
    \begin{tabular}{cccc}
        \includegraphics[width=.2\linewidth]{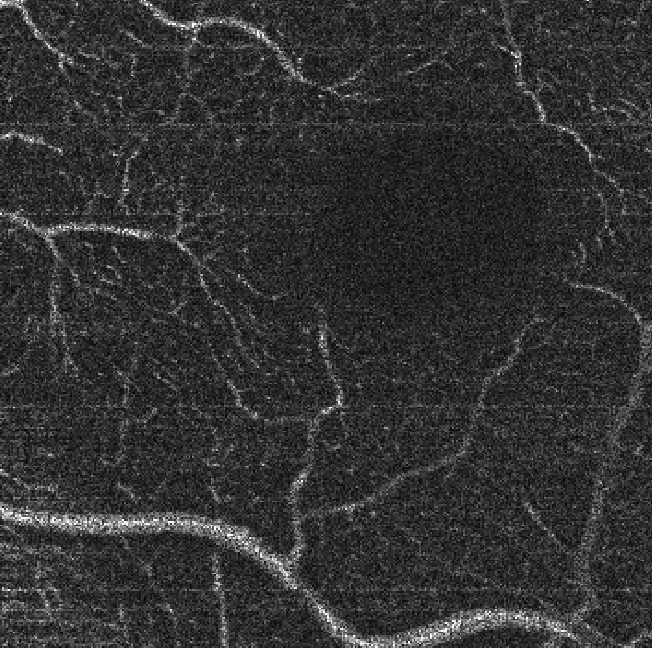} &
        \includegraphics[width=.2\linewidth]{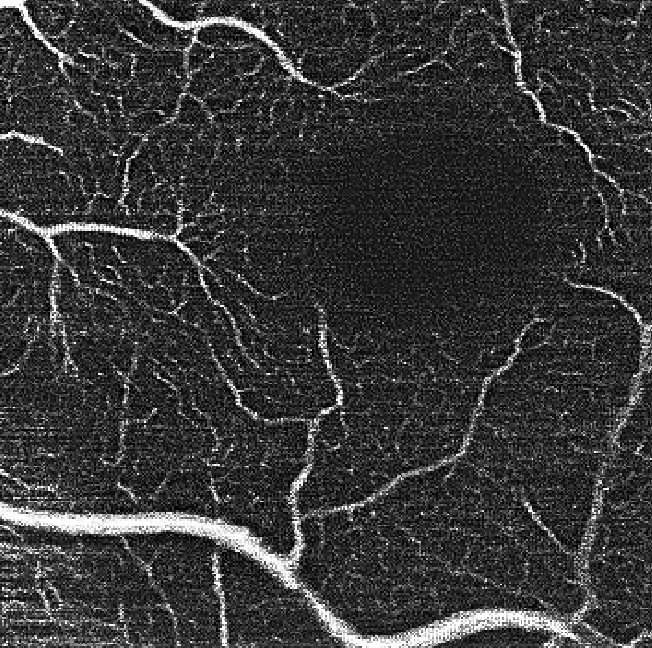} &
        \includegraphics[width=.2\linewidth]{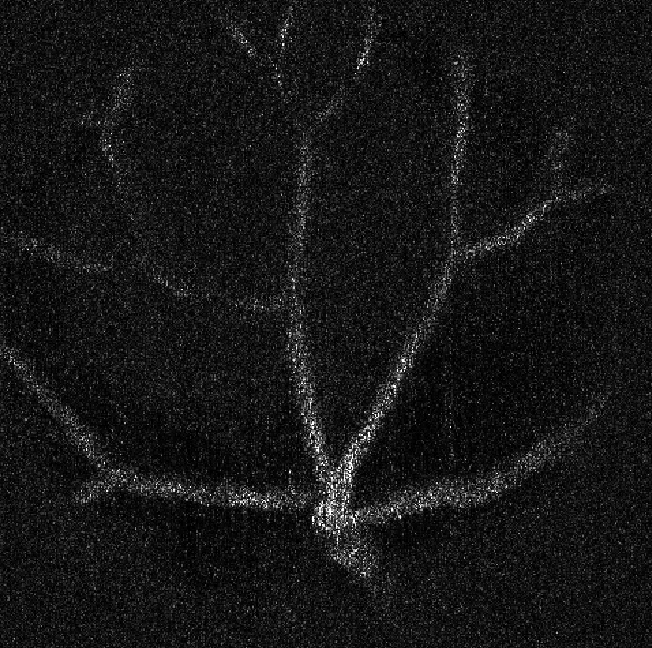} &
        \includegraphics[width=.2\linewidth]{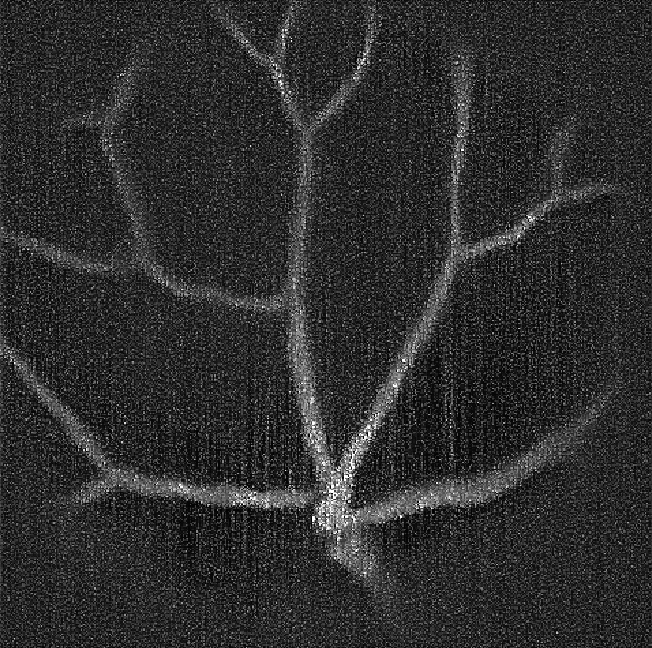} \\
        human input & human latent & fish input & fish latent
    \end{tabular}
    \caption{The latent image $L_{12}$ from LIFE considerably improves vessel appearance.}
    \label{fig:latent_example}
\end{figure}

\section{Results}
\label{sec:eval}

\begin{figure}[t]
    \centering
    \begin{tabular}{ccccccc}
    \rotatebox{90}{\hspace{0.25cm}2D slice} &
    \includegraphics[width=0.15\linewidth]{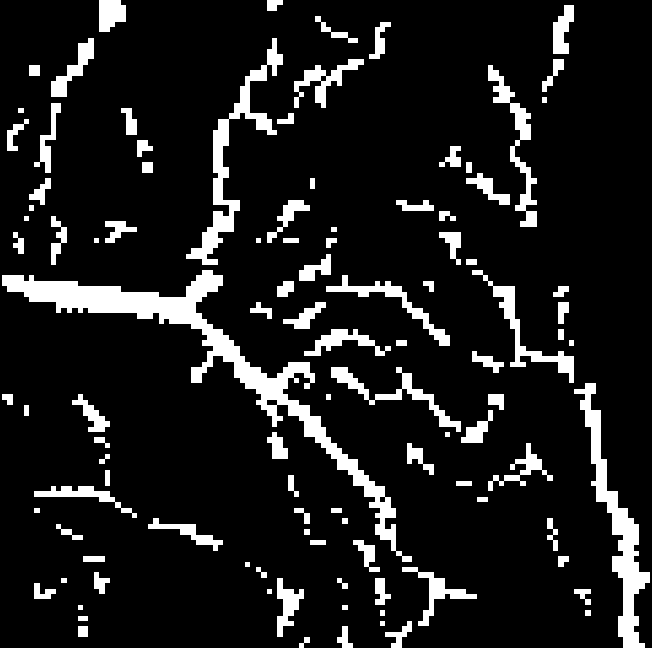} & 
    \includegraphics[width=0.15\linewidth]{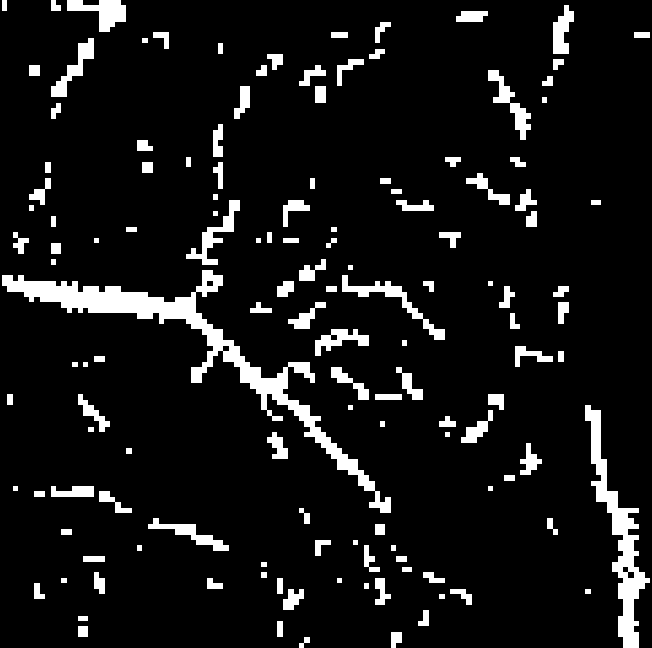} &
    \includegraphics[width=0.15\linewidth]{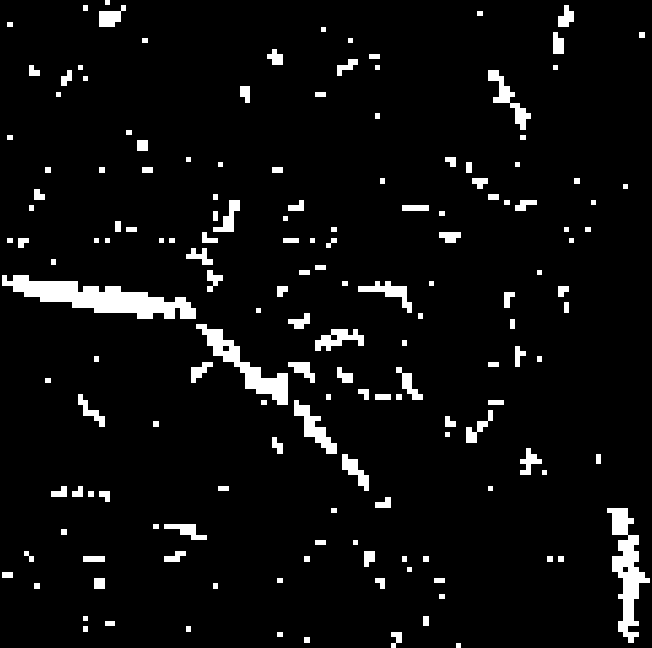} &
    \includegraphics[width=0.15\linewidth]{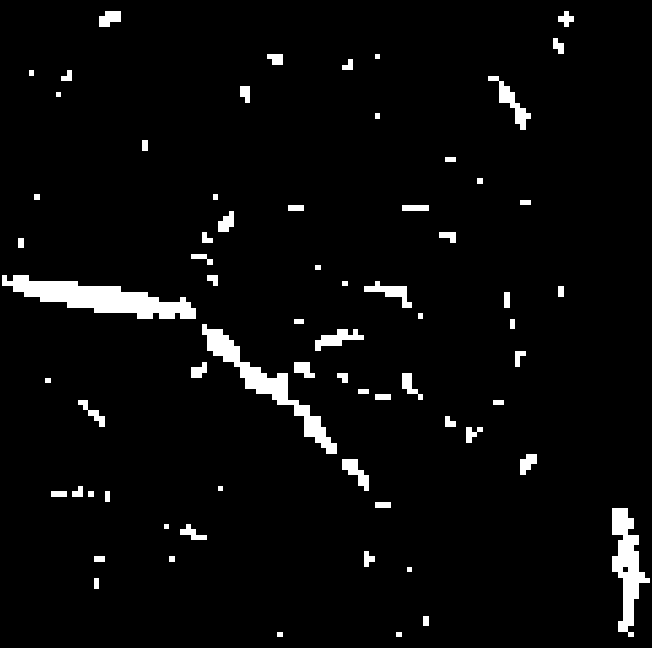} &
    \includegraphics[width=0.15\linewidth]{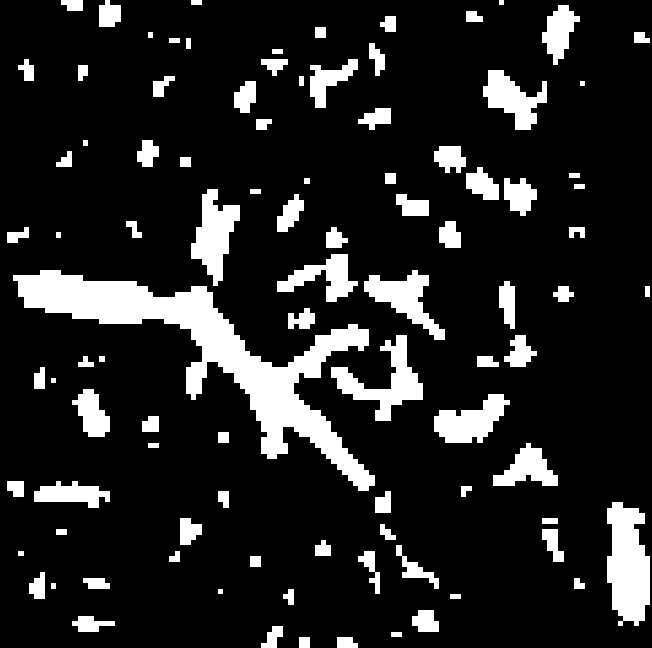} &
    \includegraphics[width=0.15\linewidth]{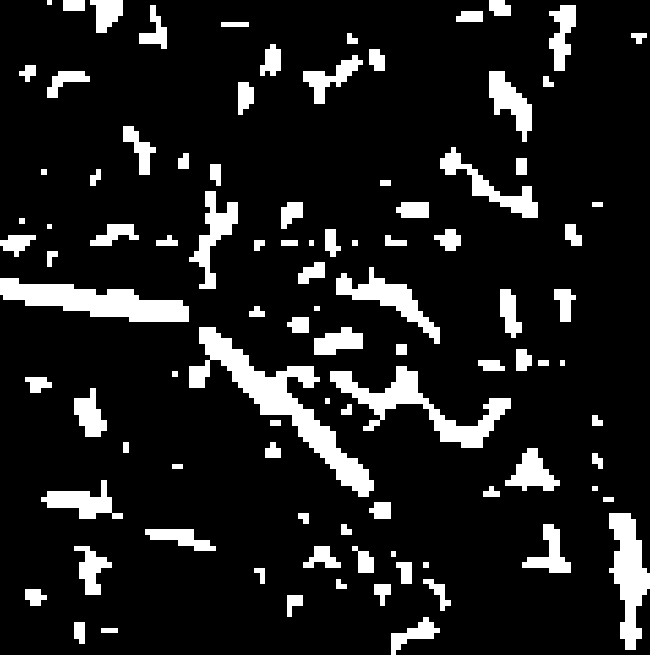}\\
    \rotatebox{90}{\hspace{0.25cm}3D, FN} &
    \includegraphics[width=0.15\linewidth]{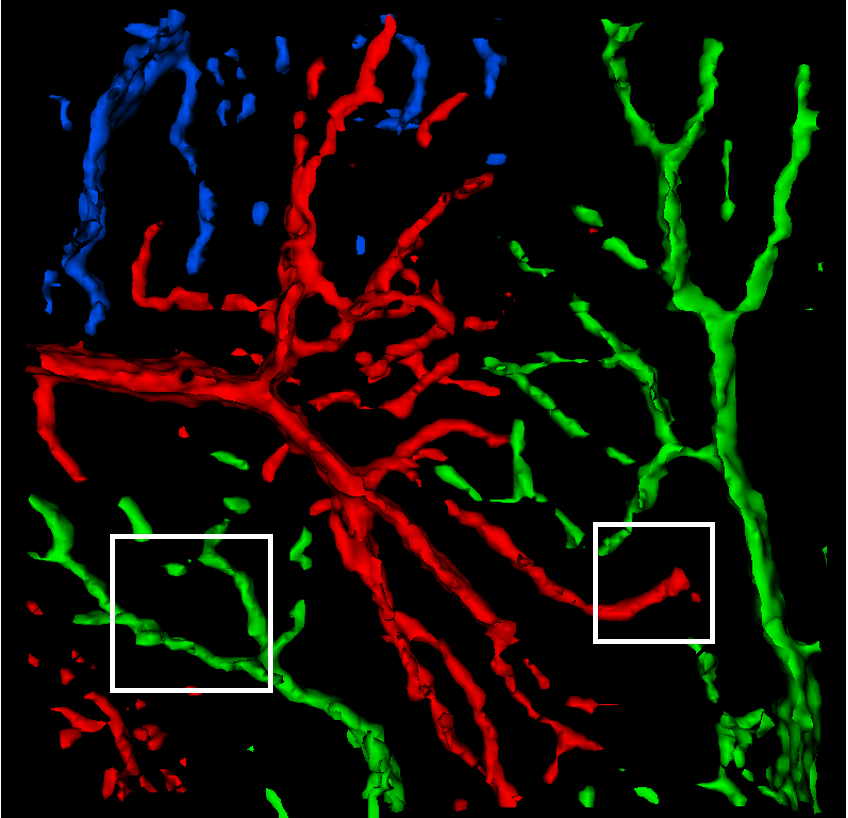} &
    \includegraphics[width=0.15\linewidth]{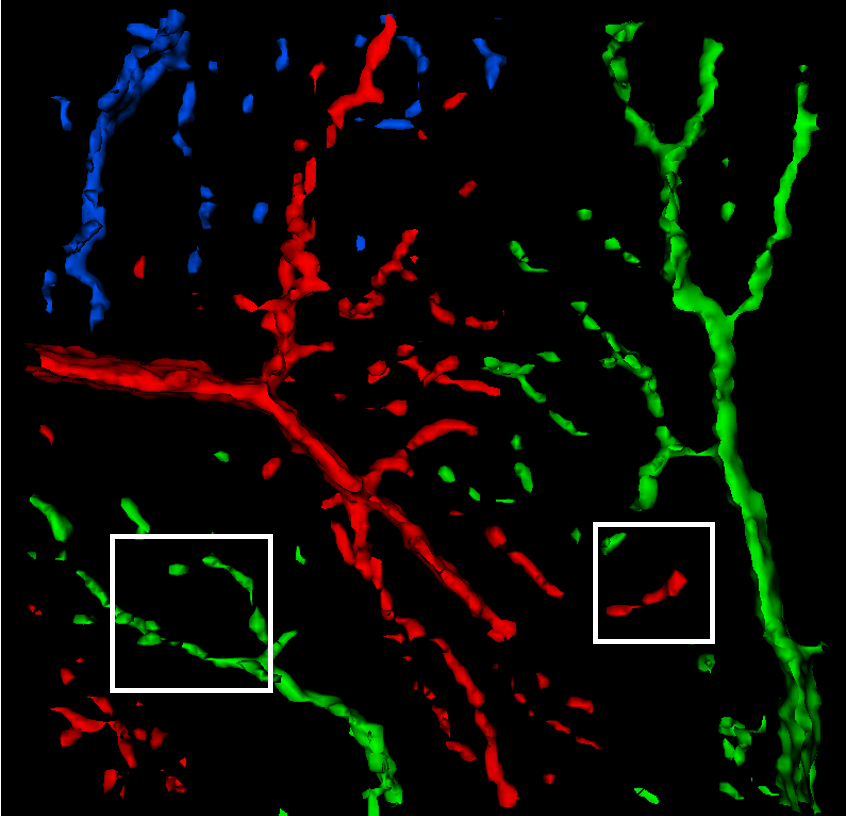} &
    \includegraphics[width=0.15\linewidth]{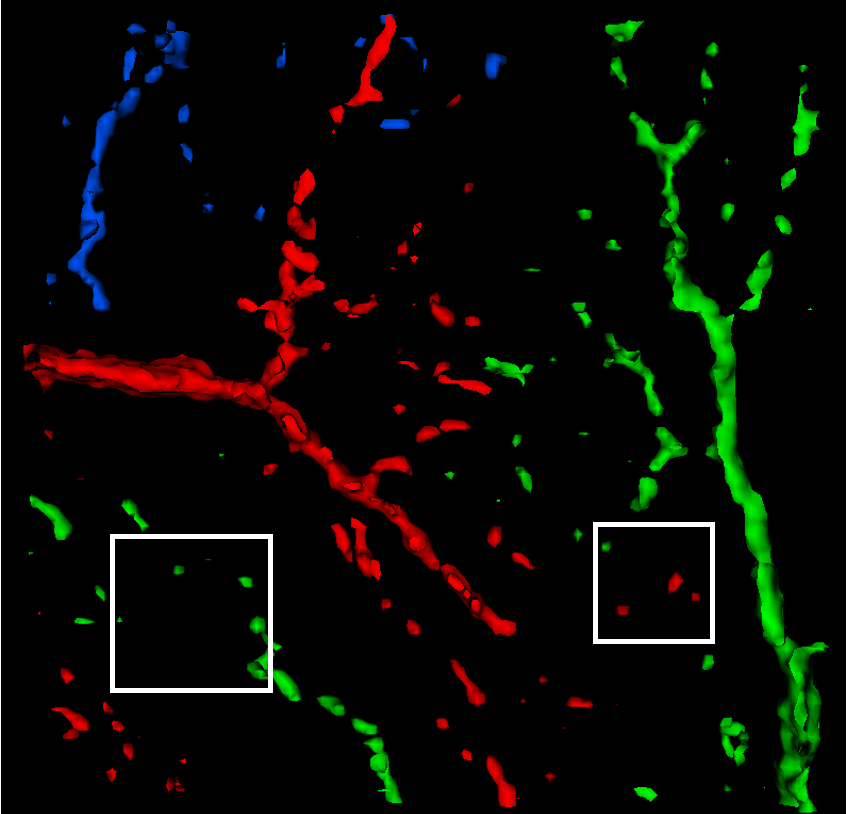} &
    \includegraphics[width=0.15\linewidth]{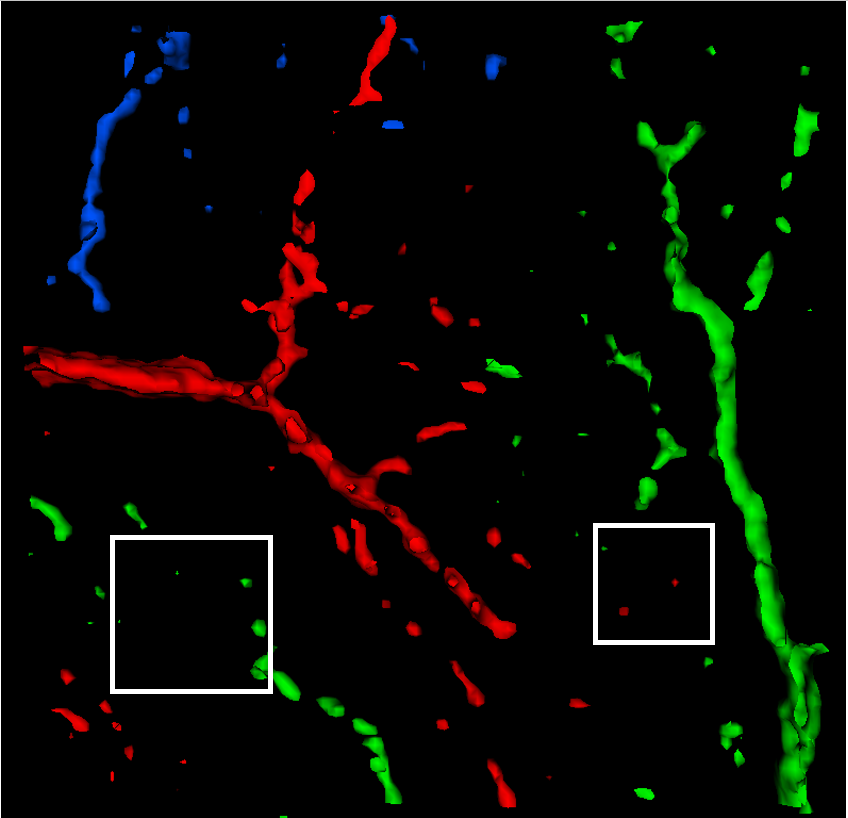} &
    \includegraphics[width=0.15\linewidth]{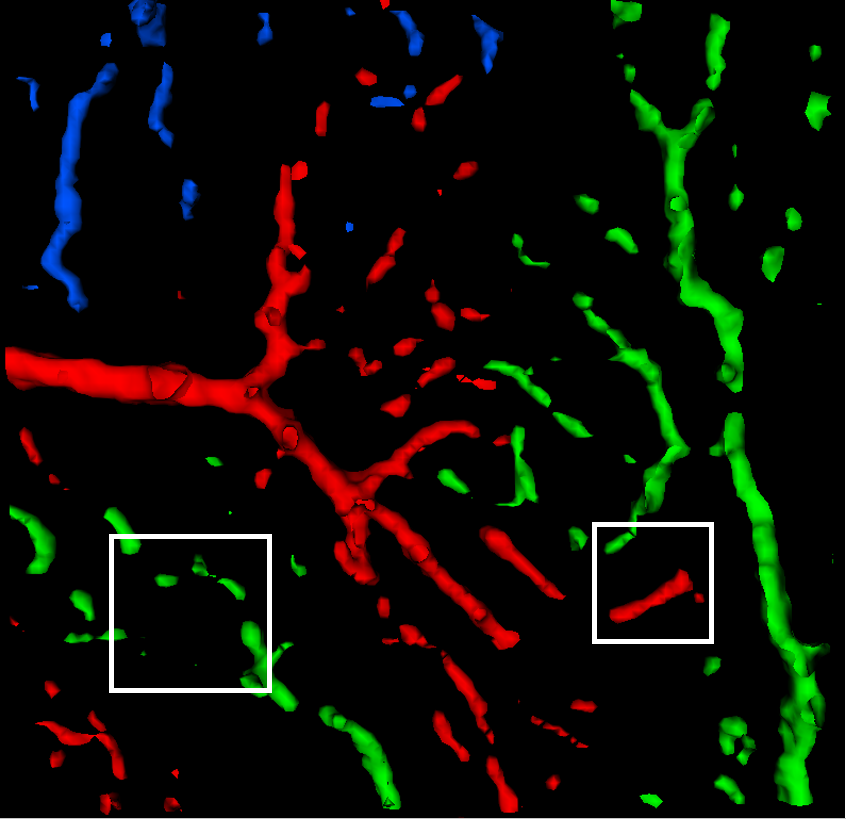} &
    \includegraphics[width=0.15\linewidth]{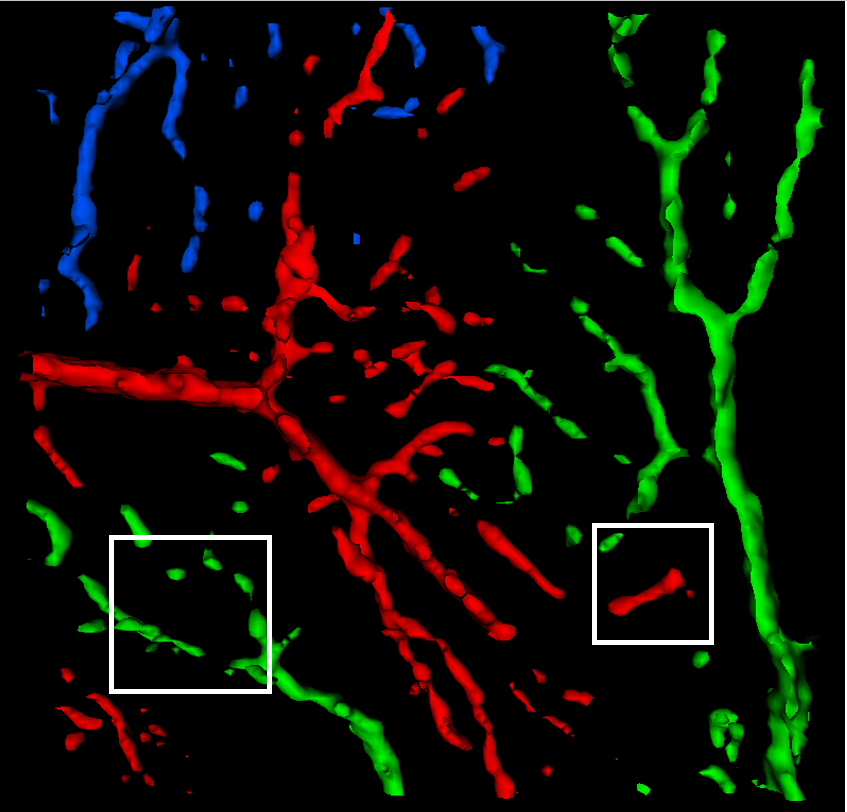} \\
    \rotatebox{90}{\hspace{0.25cm}3D, FP} &
    \includegraphics[width=0.15\linewidth]{human3d/truth_Gaussian.png} &
    \includegraphics[width=0.15\linewidth]{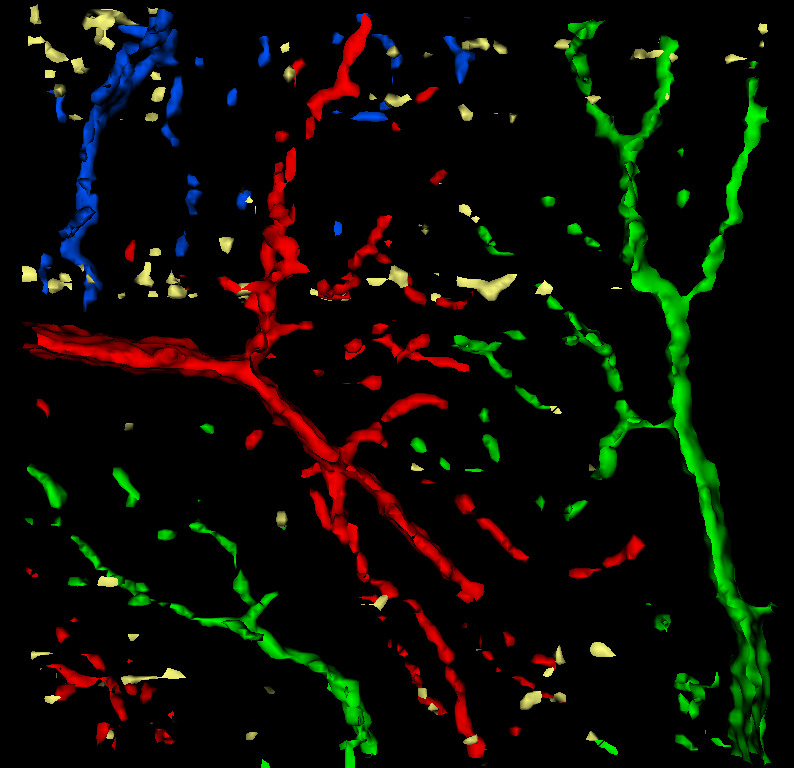} &
    \includegraphics[width=0.15\linewidth]{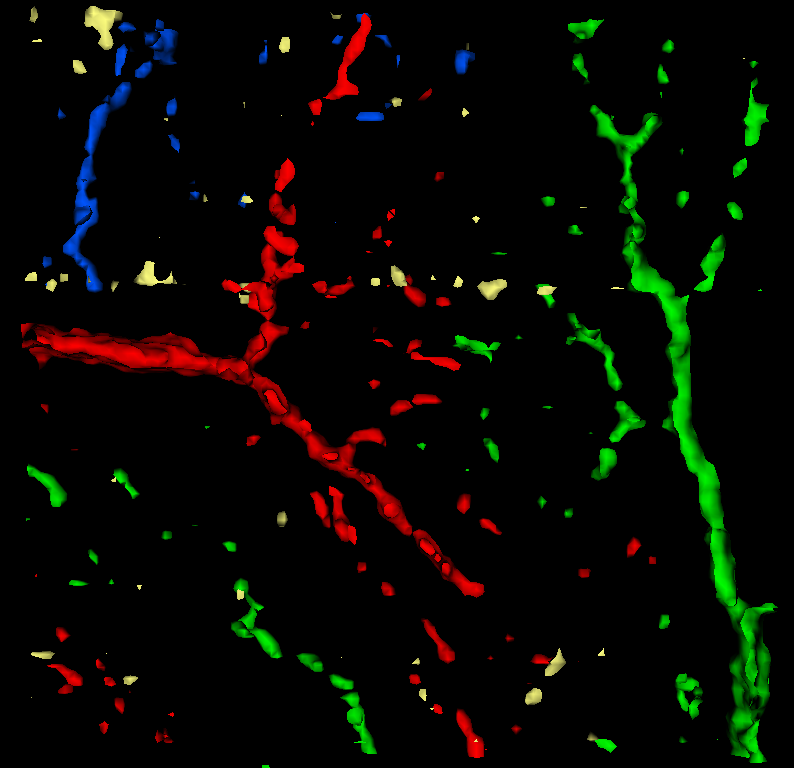} &
    \includegraphics[width=0.15\linewidth]{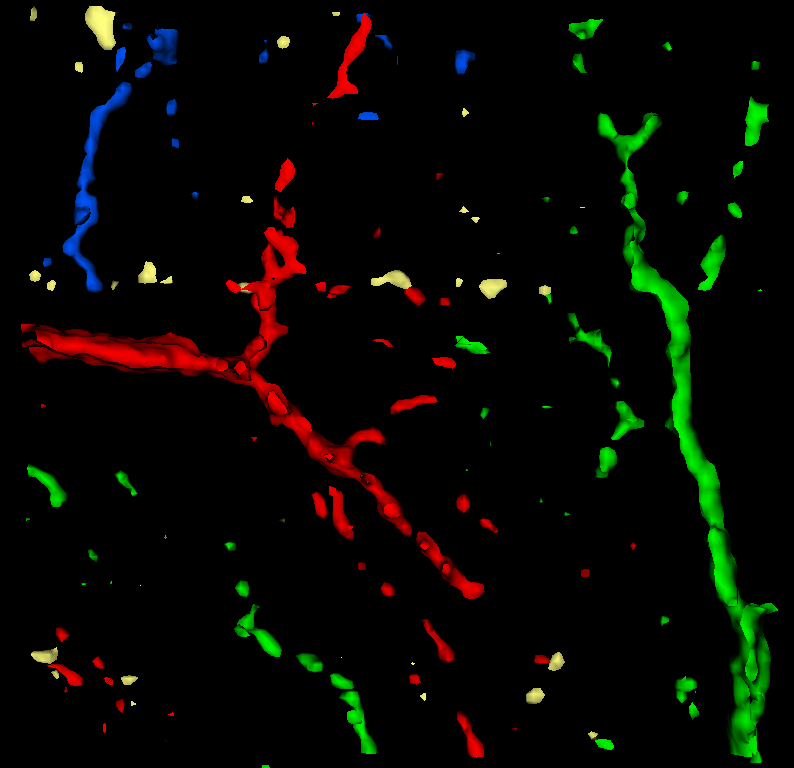} &
    \includegraphics[width=0.15\linewidth]{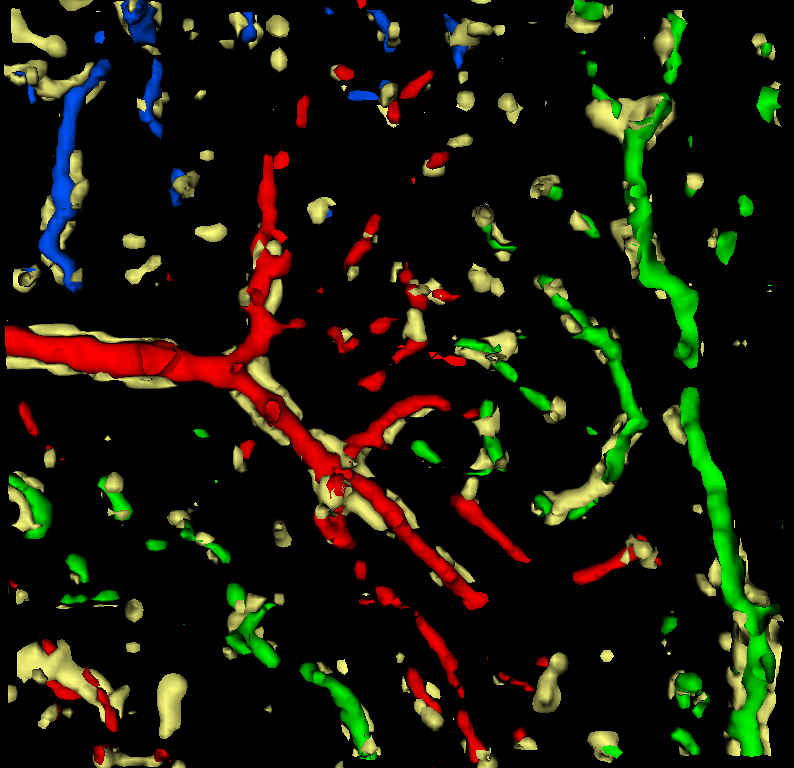} &
    \includegraphics[width=0.15\linewidth]{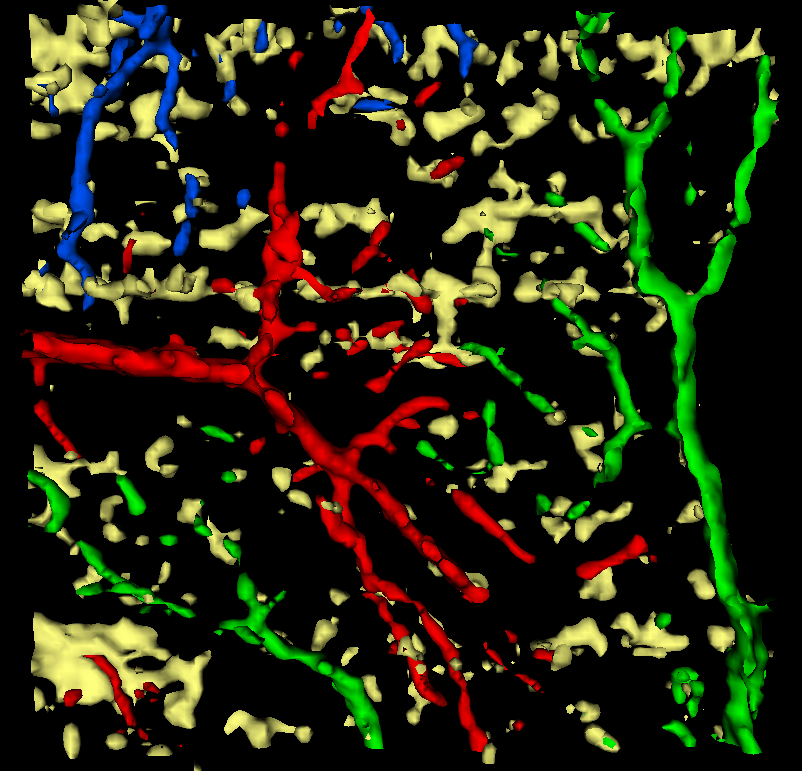}\\
    & Manual & LIFE & Otsu & K-mean & Frangi & OOF
    \end{tabular}
    \caption{2D slice and 3D marching cubes rendering of segmentation results on human retina, with Gaussian smoothing, $\sigma=0.70$. {\bf Red, green, blue} show three different branches; {\bf yellow} highlights false positives. LIFE is the only method that can recover the 3D structure and connectivity of the capillaries outside the largest vessels without causing excessive FP (yellow). {\bf White boxes} highlight LIFE's improved sensitivity. }
    \label{fig:resultrender}
\end{figure}

\begin{figure}[t]
    \centering
    \begin{tabular}{cccccc}
    \includegraphics[width=0.155\linewidth]{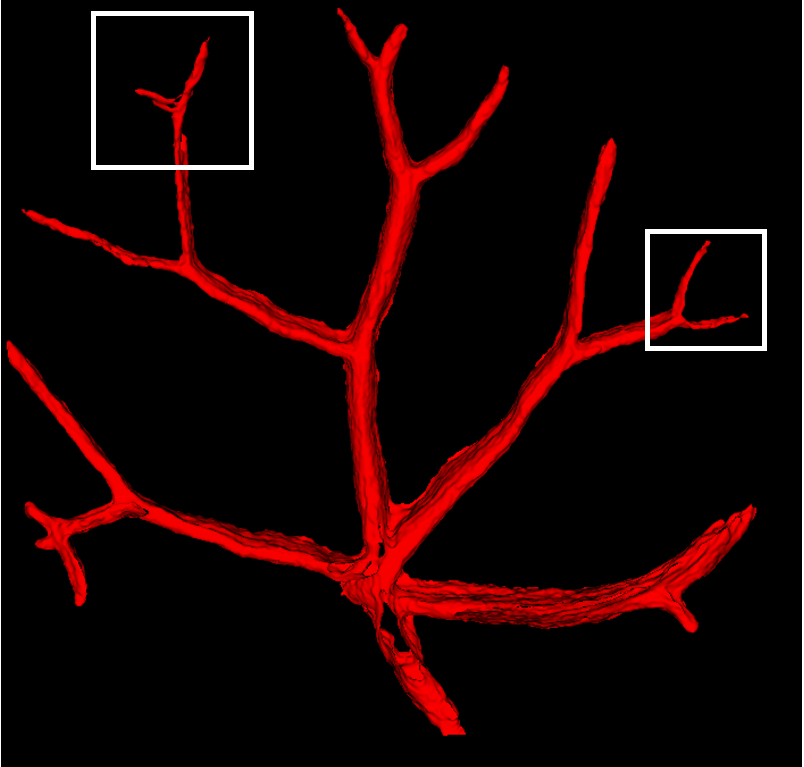} &
    \includegraphics[width=0.155\linewidth]{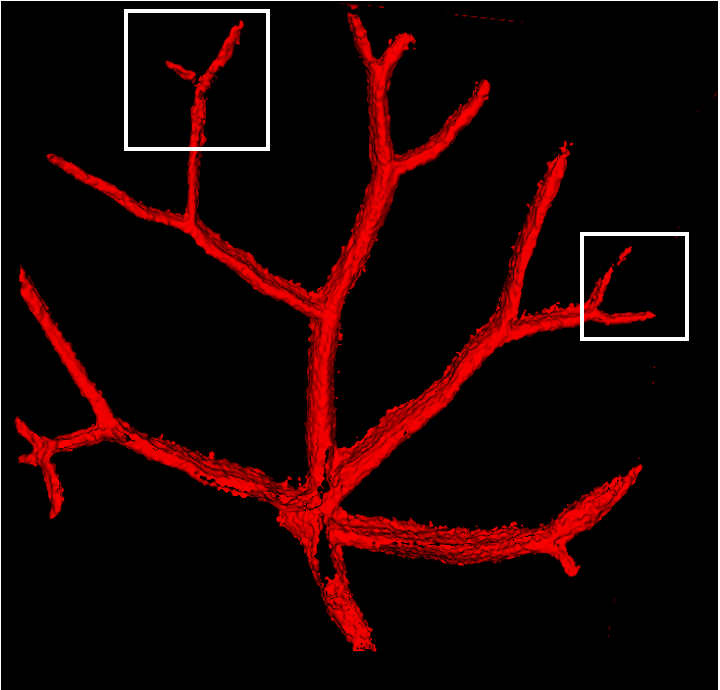} &
    \includegraphics[width=0.155\linewidth]{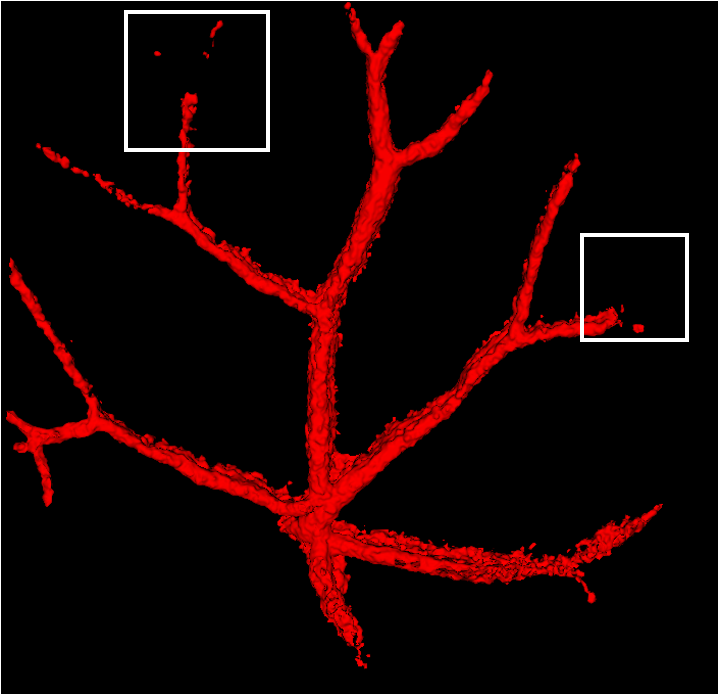} &
    \includegraphics[width=0.155\linewidth]{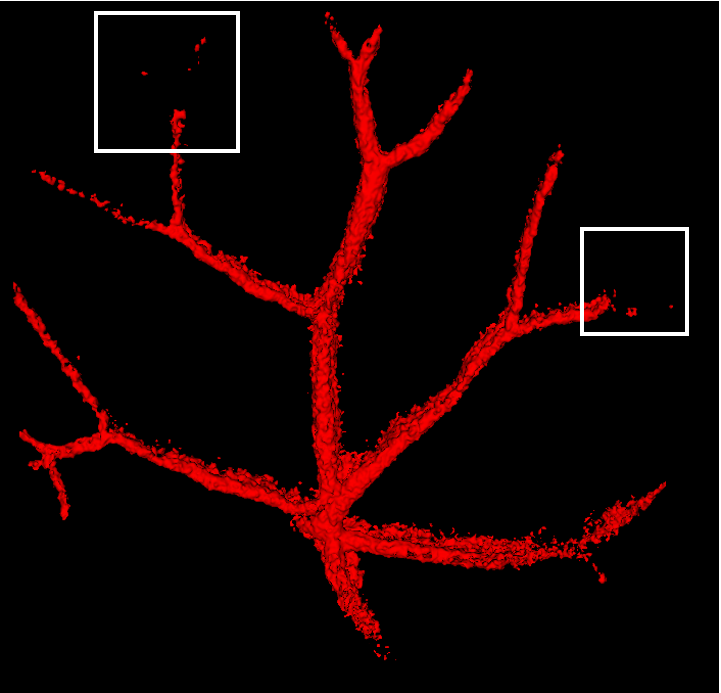} &
    \includegraphics[width=0.155\linewidth]{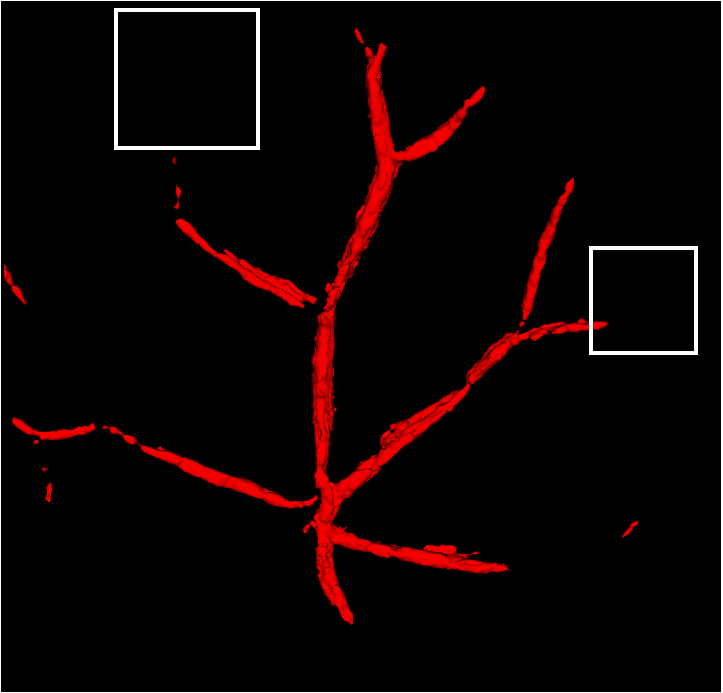} &
    \includegraphics[width=0.155\linewidth]{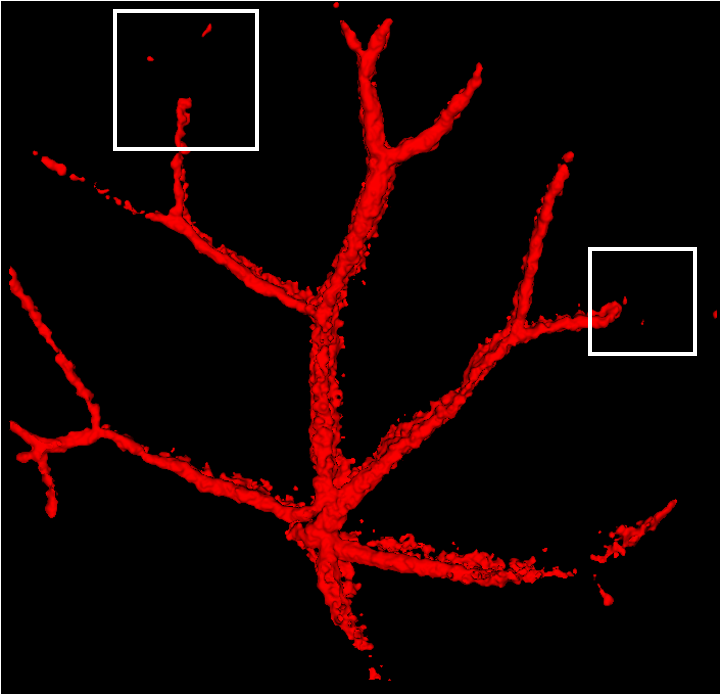} \\
    Manual & LIFE & Otsu & k-mean & Frangi & OOF 
    \end{tabular}
    \caption{segmentation result of zebrafish retina with the same rendering setting in Fig.\ref{fig:resultrender}.}
    \label{fig:fishresultrender}
\end{figure}

{\bf Fig.~\ref{fig:latent_example}} displays examples of extracted latent images. It is visually evident that LIFE successfully highlights the vasculature. Compared with the raw input, even delicate capillaries show improved homogeneity and separability from the background. {\bf Fig.~\ref{fig:resultrender}} illustrates 2D segmentation results within the manually segmented ROI, where LIFE can be seen to have better sensitivity and connectivity than the baseline methods.  Fig.~\ref{fig:resultrender} also shows a 3D rendering (via marching cubes \cite{10.1145/37402.37422}) of each method. In the middle row, we filtered out the false positives (FP) to highlight the false negatives (FN). These omitted FP areas are highlighted in yellow in the bottom row. It is easy to see that these FPs are often distributed along horizontal lines, caused by unresolved motion artifacts. Hessian-based methods appear especially sensitive to motion artifacts and noise; hence Frangi's method and OOF introduce excessive FP. Clearly, LIFE achieves the best preservation in small capillaries, such as the areas highlighted in white boxes, without introducing too many FPs.

{\bf Fig.~\ref{fig:fishresultrender}} shows that LIFE has superior performance on the zebrafish data. The white boxes highlight that only LIFE can capture smaller branches. 

\begin{figure}[b]
    \centering
    \includegraphics[width=.82\linewidth]{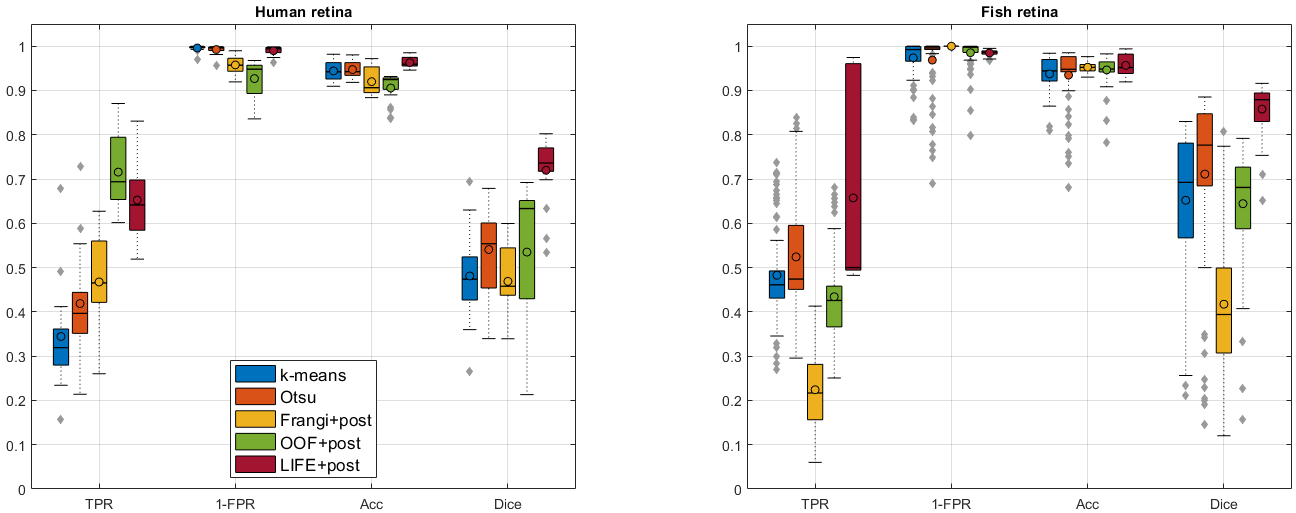}\\
    \caption{Quantitative result evaluation for  {\bf (left)} human and {\bf (right)} zebrafish data. TPR: true positive rate, FPR: false positive rate, Acc: accuracy.}
    \label{fig:boxplot}
\end{figure}

{\bf Fig.~\ref{fig:boxplot}} shows quantitative evaluation across B-scans, and {\bf Table~\ref{tab:dice}} across the whole volume. Consistent with our qualitative assessments, LIFE significantly ($p\ll 0.05$) and dramatically (over 0.20 Dice gain) outperforms the baseline methods on both human and fish data. 

Finally, we directly binarize LIF and CE-LIF as additional baselines. The Dice scores on the human data are 0.5293 and 0.4892, well below LIFE (0.7736).

\begin{table}[h]
\centering
\scalebox{1.0}{
  \begin{tabular}{l|l|l|l|l|l|l|l|l}
  \specialrule{.1em}{.05em}{.05em}
    \multirow{2}{*}{Algorithm} &
      \multicolumn{2}{c|}{TPR} &
      \multicolumn{2}{c|}{FPR} &
      \multicolumn{2}{c|}{Accuracy} &
      \multicolumn{2}{c}{Dice}\\
    & Human & Fish & Human & Fish & Human & Fish & Human & Fish \\
    \specialrule{.1em}{.05em}{.05em}
    k-means & 0.3633 & 0.4167 & $\bm{0.0042}$ & 0.0303 & 0.9440 & 0.9228 & 0.5152 & 0.6249 \\

    Otsu  & 0.4403 & 0.4356 & 0.0076 & 0.0346 & 0.9472 & 0.9191 & 0.5772 & 0.6399 \\

    Frangi+bin  & 0.4900 & 0.2117 & 0.0419 & 0.0002 & 0.9198 & $\bm{0.9489}$ & 0.5002 & 0.4212\\

    OOF+bin  & $\bm{0.6826}$ & 0.3775 & 0.0748 & 0.0165 & 0.9053 & 0.9350 & 0.5414 & 0.6247 \\

    LIFE+bin  & 0.6613 & $\bm{0.4999}$ & 0.0104 & $\bm{0.0153}$ & $\bm{0.9627}$ & 0.9386 & $\bm{0.7736}$ & $\bm{0.8594}$\\
    \specialrule{.1em}{.05em}{.05em}
  \end{tabular}}
  
  \caption{Quantitative evaluation of human and zebrafish segmentation. TPR: true positive rate, FPR: false positive rate. Bold indicates the best score per column.}
  \label{tab:dice}
\end{table}

\section{Discussion and Conclusion}
We proposed a method for 3D segmentation of fovea vessels and capillaries from OCT-A volumes that requires neither manual annotation nor multiple image acquisitions to train. The introduction of the LIF modality brings many benefits for the method. Since LIF is directly computed from the input data, no inter-volume registration is needed between the two modalities input to LIFE. Further, rather than purely depending on image intensity, LIF exploits local structural information to enhance small features like capillaries. Still, there are some disadvantages to overcome in future research. For instance, LIFE cannot directly provide a binarized output and hence the crude thresholding method used for binarization influences the segmentation performance.

\paragraph{\bf Acknowledgements.} This work is supported by NIH R01EY031769, NIH R01EY030490 and Vanderbilt University Discovery Grant Program.

\bibliographystyle{splncs04}
\bibliography{paper1556.bib}

\begin{thebibliography}{10}
\providecommand{\url}[1]{\texttt{#1}}
\providecommand{\urlprefix}{URL }
\providecommand{\doi}[1]{https://doi.org/#1}

\bibitem{alom2018recurrent}
Alom, M.Z., Hasan, M., Yakopcic, C., Taha, T.M., Asari, V.K.: {Recurrent
  residual CNN based on u-net (r2u-net) for medical image segmentation}. arXiv
  preprint arXiv:1802.06955  (2018)

\bibitem{993126}
{Aylward}, S.R., {Bullitt}, E.: Initialization, noise, singularities, and scale
  in height ridge traversal for tubular object centerline extraction. IEEE
  Transactions on Medical Imaging  \textbf{21}(2),  61--75 (2002).
  \doi{10.1109/42.993126}

\bibitem{bozkurt2020texture}
Bozkurt, F., K{\"o}se, C., Sar{\i}, A.: A texture-based 3d region growing
  approach for segmentation of ica through the skull base in cta. Multimedia
  Tools and Applications  \textbf{79}(43),  33253--33278 (2020)

\bibitem{burke2017application}
Burke, T.R., Chu, C.J., Salvatore, S., Bailey, C., Dick, A.D., Lee, R.W., Ross,
  A.H., Carre{\~n}o, E.: Application of oct-angiography to characterise the
  evolution of chorioretinal lesions in acute posterior multifocal placoid
  pigment epitheliopathy. Eye  \textbf{31}(10),  1399--1408 (2017)

\bibitem{devalla2019deep}
Devalla, S.K., Subramanian, G., Pham, T.H., Wang, X., Perera, S., Tun, T.A.,
  Aung, T., Schmetterer, L., Thi{\'e}ry, A.H., Girard, M.J.: {A deep learning
  approach to denoise OCT images of the optic nerve head}. Scientific reports
  \textbf{9}(1),  1--13 (2019)

\bibitem{el2018spectrally}
El-Haddad, M.T., Bozic, I., Tao, Y.K.: Spectrally encoded coherence tomography
  and reflectometry: Simultaneous en face and cross-sectional imaging at 2
  gigapixels per second. Journal of biophotonics  \textbf{11}(4),  e201700268
  (2018)

\bibitem{fleishman2017joint}
Fleishman, G.M., Valcarcel, A., Pham, D.L., Roy, S., Calabresi, P.A.,
  Yushkevich, P., Shinohara, R.T., Oguz, I.: {Joint intensity fusion image
  synthesis applied to MS lesion segmentation}. In: MICCAI BrainLes Workshop.
  pp. 43--54 (2017)

\bibitem{frangi1998multiscale}
Frangi, A.F., Niessen, W.J., Vincken, K.L., Viergever, M.A.: Multiscale vessel
  enhancement filtering. In: {MICCAI}. pp. 130--137. Springer (1998)

\bibitem{gao2016optical}
Gao, S., Jia, Y., Zhang, M., Su, J., Liu, G., Hwang, T., Bailey, S., Huang, D.:
  Optical coherence tomography angiography. IOVS  \textbf{57}(9),  OCT27--OCT36
  (2016)

\bibitem{giarratano2019automated}
Giarratano, Y., Bianchi, E., Gray, C., Morris, A., MacGillivray, T., Dhillon,
  B., Bernabeu, M.O.: Automated and network structure preserving segmentation
  of optical coherence tomography angiograms. arXiv preprint arXiv:1912.09978
  (2019)

\bibitem{guizar2008efficient}
Guizar-Sicairos, M., Thurman, S.T., Fienup, J.R.: Efficient subpixel image
  registration algorithms. Optics letters  \textbf{33}(2),  156--158 (2008)

\bibitem{hollo2018comparison}
Holl{\'o}, G.: Comparison of peripapillary oct angiography vessel density and
  retinal nerve fiber layer thickness measurements for their ability to detect
  progression in glaucoma. Journal of glaucoma  \textbf{27}(3),  302--305
  (2018)

\bibitem{hu2020retinal}
Hu, D., Malone, J., Atay, Y., Tao, Y., Oguz, I.: {Retinal OCT Denoising with
  Pseudo-Multimodal Fusion Network}. In: MICCAI OMIA. pp. 125--135 (2020)

\bibitem{ishibazawa2015optical}
Ishibazawa, A., Nagaoka, T., Takahashi, A., Omae, T., Tani, T., Sogawa, K.,
  Yokota, H., Yoshida, A.: {OCT} angiography in diabetic retinopathy: a
  prospective pilot study. American journal of ophthalmology  \textbf{160}(1),
  35--44 (2015)

\bibitem{jia2012split}
Jia, Y., Tan, O., Tokayer, J., Potsaid, B., Wang, Y., Liu, J.J., Kraus, M.F.,
  Subhash, H., Fujimoto, J.G., Hornegger, J., et~al.: Split-spectrum
  amplitude-decorrelation angiography with optical coherence tomography. Optics
  express  \textbf{20}(4),  4710--4725 (2012)

\bibitem{kingma2013auto}
Kingma, D.P., Welling, M.: Auto-encoding variational bayes. arXiv preprint
  arXiv:1312.6114  (2013)

\bibitem{lahiri2016deep}
Lahiri, A., Roy, A.G., Sheet, D., Biswas, P.K.: Deep neural ensemble for
  retinal vessel segmentation in fundus images towards achieving label-free
  angiography. In: IEEE EMBC. pp. 1340--1343. IEEE (2016)

\bibitem{law2008three}
Law, M.W., Chung, A.C.: Three dimensional curvilinear structure detection using
  optimally oriented flux. In: European conference on computer vision. pp.
  368--382. Springer (2008)

\bibitem{li2017statistical}
Li, M., Idoughi, R., Choudhury, B., Heidrich, W.: Statistical model for oct
  image denoising. Biomedical Optics Express  \textbf{8}(9),  3903--3917 (2017)

\bibitem{liu2020variational}
Liu, Y., Zuo, L., Carass, A., He, Y., Filippatou, A., Solomon, S.D., Saidha,
  S., Calabresi, P.A., Prince, J.L.: Variational intensity cross channel
  encoder for unsupervised vessel segmentation on oct angiography. In: SPIE
  Medical Imaging 2020: Image Processing. vol. 11313, p. 113130Y (2020)

\bibitem{10.1145/37402.37422}
Lorensen, W.E., Cline, H.E.: Marching cubes: A high resolution 3d surface
  construction algorithm. SIGGRAPH Comput. Graph.  \textbf{21}(4),  163–169
  (Aug 1987). \doi{10.1145/37402.37422},
  \url{https://doi.org/10.1145/37402.37422}

\bibitem{Lorigo:2001jv}
Lorigo, L.M., Faugeras, O.D., Grimson, W.E., Keriven, R., Kikinis, R., Nabavi,
  A., Westin, C.F.: {CURVES: curve evolution for vessel segmentation.} Medical
  Image Analysis  \textbf{5}(3),  195--206 (Sep 2001)

\bibitem{malone2019handheld}
Malone, J.D., El-Haddad, M.T., Yerramreddy, S.S., Oguz, I., Tao, Y.K.:
  {Handheld spectrally encoded coherence tomography and reflectometry for
  motion-corrected ophthalmic OCT and OCT-A}. Neurophotonics  \textbf{6}(4),
  041102 (2019)

\bibitem{oguz2020self}
Oguz, I., Malone, J.D., Atay, Y., Tao, Y.K.: {Self-fusion for OCT noise
  reduction}. In: SPIE Medical Imaging 2020: Image Processing. vol. 11313, p.
  113130C (2020)

\bibitem{otsu1979threshold}
Otsu, N.: A threshold selection method from gray-level histograms. IEEE
  transactions on systems, man, and cybernetics  \textbf{9}(1),  62--66 (1979)

\bibitem{perona1990scale}
Perona, P., Malik, J.: Scale-space and edge detection using anisotropic
  diffusion. IEEE Trans.~on pattern analysis and machine intelligence
  \textbf{12}(7),  629--639 (1990)

\bibitem{10.1117/12.2550009}
Ufford, K., Vandekar, S., Oguz, I.: {Joint intensity fusion with normalized
  cross-correlation metric for cross-modality MRI synthesis}. In: SPIE Medical
  Imaging 2020: Image Processing. vol. 11313 (2020)

\bibitem{Vasilevskiy01fluxmaximizing}
Vasilevskiy, A., Siddiqi, K.: Flux maximizing geometric flows. IEEE
  Transactions on Pattern Analysis and Machine Intelligence  \textbf{24},
  1565--1578 (2001)

\bibitem{wang2012multi}
Wang, H., Suh, J.W., Das, S.R., Pluta, J.B., Craige, C., Yushkevich, P.A.:
  Multi-atlas segmentation with joint label fusion. IEEE PAMI  \textbf{35}(3),
  611--623 (2012)

\bibitem{py06nimg}
Yushkevich, P.A., Piven, J., Cody~Hazlett, H., Gimpel~Smith, R., Ho, S., Gee,
  J.C., Gerig, G.: User-guided {3D} active contour segmentation of anatomical
  structures: Significantly improved efficiency and reliability. Neuroimage
  \textbf{31}(3),  1116--1128 (2006)

\bibitem{yushkevich2016fast}
Yushkevich, P.A., Pluta, J., Wang, H., Wisse, L.E., Das, S., Wolk, D.: {Fast
  automatic segmentation of hippocampal subfields and medial temporal lobe
  subregions in 3T and 7T T2-weighted MRI}. Alzheimer's \& Dementia
  \textbf{7}(12),  P126--P127 (2016)

\bibitem{8879535}
{Zhang}, J., {Qiao}, Y., {Sarabi}, M.S., {Khansari}, M.M., {Gahm}, J.K.,
  {Kashani}, A.H., {Shi}, Y.: 3d shape modeling and analysis of retinal
  microvasculature in oct-angiography images. IEEE TMI  \textbf{39}(5),
  1335--1346 (2020)

\bibitem{zhao2018vascular}
Zhao, S., Tian, Y., Wang, X., Xu, P., Deng, Q., Zhou, M.: Vascular extraction
  using mra statistics and gradient information. Mathematical Problems in
  Engineering  \textbf{2018} (2018)

\end{thebibliography}

\clearpage

\end{document}